\documentclass[fleqn,usenatbib]{mnras}

\usepackage{newtxtext,newtxmath}
\usepackage[T1]{fontenc}

\DeclareRobustCommand{\VAN}[3]{#2}
\let\VANthebibliography\thebibliography
\def\thebibliography{\DeclareRobustCommand{\VAN}[3]{##3}\VANthebibliography}

\usepackage{graphicx}	
\usepackage{amsmath}	
\usepackage{wasysym}           
\usepackage{graphicx}
\usepackage{epstopdf}
\usepackage{mathrsfs}
\usepackage{anyfontsize}
\usepackage{natbib}
\usepackage{color}
\usepackage{lipsum}
\usepackage{diagbox}

\title[The dynamics of tidally disrupted debris streams]{The dynamics of debris streams from tidal disruption events: exact solutions, critical stream density, and hydrogen recombination}

\author[E.~R.~Coughlin]{
Eric R.~Coughlin$^{1}$\thanks{E-mail: ecoughli@syr.edu}
\\
$^{1}$Department of Physics, Syracuse University, Syracuse, NY 13244, USA
}

\date{Accepted XXX. Received YYY; in original form ZZZ}

\pubyear{2020}

\begin{document}
\label{firstpage}
\pagerange{\pageref{firstpage}--\pageref{lastpage}}
\maketitle

\begin{abstract}
A star destroyed by a supermassive black hole (SMBH) in a tidal disruption event (TDE) is transformed into a filamentary structure known as a tidally disrupted stellar debris stream. We show that when ideal gas pressure dominates the thermodynamics of the stream, there is an exact solution to the hydrodynamics equations that describes the stream evolution and accounts for self-gravity, pressure, the dynamical expansion of the gas, and the transverse structure of the stream. We analyze the stability of this solution to cylindrically symmetric perturbations, and show that there is a critical stream density below which the stream is unstable and is not self-gravitating; this critical density is a factor of at least 40-50 smaller than the stream density in a TDE. Above this critical density the stream is overstable, self-gravity confines the stream, the oscillation period is exponentially long, and the growth rate of the overstability scales as $t^{1/6}$. The power-law growth and small power-law index of the overstability implies that the stream is effectively stable to cylindrically symmetric perturbations. We also use this solution to analyze the effects of hydrogen recombination, and suggest that even though recombination substantially increases the gas entropy, it is likely incapable of completely destroying the influence of self-gravity. We also show that the transient produced by recombination is far less luminous than previous estimates. 
\end{abstract}


\section{Introduction}
The destruction of a star by the tides of a supermassive black hole (SMBH), known as a tidal disruption event (TDE; e.g., \citealt{rees88, gezari21}), has garnered significant attention over the last decade with the advent of survey science. Specifically, astronomical surveys such as the All-Sky Automated Search for SuperNovae (ASAS-SN; \citealt{shappee14}), the Palomar Transient Factory (PTF; \citealt{law19}), the Panoramic Survey Telescope and Rapid Response System (Pan-STARRs; \citealt{kaiser10}), the Asteroid Terrestrial Last Alert Survey (ATLAS; \citealt{tonry18}), and the Zwicky Transient Facility (ZTF; \citealt{bellm19}) have been discovering TDEs at an extremely elevated rate, with $\gtrsim 100$ plausible events now detected (see \citealt{gezari21} for a review of the observational status). In the forthcoming years, the Legacy Survey of Space and Time/Rubin Observatory (LSST; \citealt{ivezic19}) is expected to increase the number of observed TDEs by at least a factor of $\sim few-10$, and potentially larger still \citep{bricman20}. 

From a theoretical standpoint, the initial phases of what one might call the ``canonical TDE'' -- a $5/3$ polytrope with a radius $R_{\star} = 1R_{\odot}$ and mass $M_{\star} = 1M_{\odot}$ destroyed by a SMBH of mass $M_{\bullet} = 10^6M_{\odot}$ as it passes through the tidal radius $r_{\rm t} = R_{\star}\left(M_{\bullet}/M_{\star}\right)^{1/3}$ -- is well understood. In particular, the specific energy spread imparted to the tidal debris as the star is destroyed, as described by \citet{lacy82} (and $\sim 100$ times larger than the binding energy of the original star), was recovered numerically as early as \citet{evans89}; these authors also reproduced the $\propto t^{-5/3}$ ``fallback rate,'' or the rate at which tidally destroyed material returns to the SMBH, predicted by \citet{rees88,phinney89}. The destruction of the polytrope itself was modeled even earlier with smoothed-particle hydrodynamics (SPH) by \citet{nolthenius82} using 40 particles, showing the formation of an elongated tendril of gas -- hereafter referred to as a tidally disrupted stellar debris stream. Since then, the {initial destruction} of the star during the canonical TDE, i.e., for the first $\sim 100$ dynamical times of the initial star after its pericenter passage, has been modeled with both increasing levels of accuracy in SPH (both in terms of numerical technique, i.e., variable smoothing lengths and artificial viscosity, and particle number) and finite volume methods (e.g., \citealt{laguna93, ayal00, lodato09, guillochon13, mainetti17}). 

The earliest work (of which we are aware) on the \emph{long-term} evolution of the tidally disrupted debris stream produced from a TDE, which amounts to discerning the ultimate fate of the debris stream $\gtrsim 1000$s of dynamical times after the initial encounter (and longer), was performed by \citet{kochanek94}. \citet{kochanek94} developed a ``one-zone'' model, in which the stream is modeled in a Lagrangian sense and broken into a number of segments, each of which has a length and cross-sectional width $H$, across which the density does not vary within the model. The length of each segment is established by assuming ballistic motion in the gravitational field of the SMBH, with leading-order general relativistic effects included, while the transverse dynamics are constrained by self-gravity, pressure, and the tidal force of the SMBH. \citet{kochanek94} also included the effects of viscosity, both shear and bulk, and analyzed the effects of recombination and the AGN luminosity of the SMBH (should the SMBH be in an actively accreting state) on the stream structure. He then investigated the nature of the caustics by following the evolution of the stream back to pericenter and -- since he accounted for general relativistic apsidal precession -- self-intersection, and also concluded that Lense-Thirring (i.e., nodal) precession could delay the self-intersections that are otherwise induced by apsidal precession. \citet{kochanek94} concluded that the evolution of the debris stream is ``complicated and involves self-gravity, tidal gravity, recombination, \ldots'' 

Only relatively recently has the problem of the long-term debris stream evolution been revisited. Specifically, \citet{kasen10} investigated the possibility of a ``recombination transient'' that could occur once the stream started to recombine. Unlike \citet{kochanek94}, they let the debris stream expand ballistically in all directions and did not account for the self-gravitating nature of the stream in the transverse directions.  \citet{guillochon14} numerically studied the debris stream produced from a TDE with the finite-volume code {\sc flash} \citep{fryxell00} and found good agreement with the predictions of \citet{kochanek94}. \citet{coughlin16} developed a semi-analytical model for the evolution of the debris stream, which differed from the work of \citet{kochanek94} in that it was Eulerian and used a self-similar solution for the expansion of the gas in the radial direction. \citet{guillochon16} used a Lagrangian approach to understand the effects of drag from the circumnuclear medium on the propagation of the unbound debris from a TDE, and \citet{bonnerot16} used a Lagrangian method to understand the effects of the ambient medium on the stream (and in particular the Kelvin-Helmholtz instability). Most recently \citet{bonnerot22} used a Lagrangian technique to model the evolution of the debris stream from the initial destruction of the star to the return of the most bound debris, using the frozen-in approximation for the initial conditions (\citealt{kochanek94} used the affine-star model of \citealt{carter83} to establish the initial conditions to solve his equations of motion).

\citet{coughlin15}, who numerically simulated the debris stream evolution from the canonical TDE with the SPH code {\sc phantom} \citep{price18}, found that the stream would fragment under its own self-gravity into localized knots (see also \citealt{hayasaki20, sacchi20}), which was not predicted by earlier models (\citealt{guillochon14a} argued that the combination of radiative cooling and Kelvin-Helmholtz instability could result in the formation of a clump, and did not account for the stream self-gravity in their simulations). Using their Eulerian and semi-analytical model described above, \citet{coughlin16} argued that the ability of the stream to fragment under its own self-gravity is critically related to its equation of state. Specifically, they showed that equations of state stiffer (softer) than a $\gamma = 5/3$ adiabatic equation of state are unstable (stable) to fragmentation near the marginally bound radius of the stream (as further supported by the numerical investigations in \citealt{coughlin16b}). They also found that different parts of the stream in terms of their Keplerian binding energy to the SMBH were more or less susceptible to fragmentation, with unbound portions of the stream (which eventually transition to homologous expansion) being gravitationally unstable with an equation of state as soft as $\gamma = 4/3$. Despite their arguments and numerical investigations, a number of questions regarding the stability of the stream and its self-gravitating nature remain, such as the necessary conditions for it to be self-gravitating in the first place. 

It is the purpose of the present investigation to analytically (although we also make use of numerical simulations to substantiate our conclusions; see Figures \ref{fig:rho_of_r}, \ref{fig:debris_stream}, and \ref{fig:rho_rhobullet} below) and quantitatively understand the stability and the self-gravitating nature of the debris streams produced from TDEs, and in particular the role played by the transverse structure of the stream. After discussing some basic considerations of the problem in Section \ref{sec:basic} and the coordinate system we adopt in Section \ref{sec:coordinates}, in Section \ref{sec:exact} we show that there is an exact solution to the fluid equations that describes a self-gravitating, pressurized fluid that is expanding both along the axis of the filament and perpendicularly to it (and at different rates), the transverse structure of which is determined by the equation of hydrostatic balance (despite the fact that the system is not actually hydrostatic) and the Poisson equation. This solution is always valid near the marginally bound Keplerian radius, and holds over the entire stream until the most-bound material (to the SMBH) reaches its apocenter, which is $\sim 1000$'s of dynamical times of the star for typical SMBH masses. We identify this solution as the background state of the stream. 

In Section \ref{sec:perturbations} we analyze cylindrically symmetric perturbations on top of this background solution, and show that there is a critical value of the stream density that divides purely unstable and overstable oscillations of the stream. We identify this critical density as the one necessary for the stream to remain self-gravitating, and it is $\sim 1-2$ orders of magnitude below what is achieved in a TDE (see Section \ref{sec:basic}), suggesting that effectively all TDE streams are confined by self-gravity. We also show that the growth of the overstability is $\propto t^{1/6}$, which is extremely weakly growing, and the oscillation period of the overstability is exponentially long. We investigate the effects of recombination in Section \ref{sec:recombination}, and suggest that while it substantially modifies the stream thermodynamics and sets in after $\sim 80$ dynamical times of the original star (corresponding to $\sim 1.5$ days for a star of one solar mass and radius), it likely cannot completely destroy the influence of self-gravity because of the large difference between the critical density needed to be self-gravitating and the (much larger) density of the stream. We summarize and conclude in Section \ref{sec:summary}. In Appendix \ref{sec:homologous} we also show that a close analog of the one-zone model of \citet{kochanek94} can be rigorously derived from the fluid equations, and is the leading-order term in a series expansion of the Lagrangian position of a fluid element in terms of its initial position.  

\section{Basic Considerations}
\label{sec:basic}
As the star passes through the tidal disruption radius, the usual assumption that the fluid moves predominantly with the center of mass and thereafter evolves quasi-ballistically in the gravitational field of the black hole \citep{lacy82} implies that the Keplerian energies of fluid elements are -- to leading order in the ratio of the stellar radius to the tidal radius -- functions only of the cylindrical distance from the black hole \citep{lodato09}. Therefore, even if the tidally disrupted material were assumed to continue evolving purely ballistically, it would do so in the form of a filamentary structure, the cross-sectional radius of which is much less than its radial extent (e.g., Figure 1 of \citealt{kasen10}, or Figure 2 of \citealt{coughlin16b}). The question then becomes -- is it valid to ignore self-gravity completely as concerns the evolution of the stream? 

The relevance of self-gravity can be understood by considering the ratio of the stream density $\rho_{\rm i}$ to the ``black hole density,'' where the latter is defined as $\rho_{\bullet} = M_{\bullet}/(4\pi r^3/3)$ with $M_{\bullet}$ the mass of the SMBH and $r$ is the Lagrangian position of a fluid element within the stream; if this ratio is comparable to one, self-gravity is important for the dynamics of the gas (e.g., \citealt{pringle81} and references therein in the context of accretion discs). This condition arises from the fact that the vertical component of the gravitational field of the SMBH is $\sim GMH/r^3 \simeq G\rho_{\bullet}H$ if the cross-sectional radius of the stream is $H$ (see Equation \ref{smom1} below), while the self-gravitational field is $\sim G\rho_{\rm i} H$ (Equations \ref{smom1} and \ref{phi1} below). \citet{coughlin22a} pointed out that an accurate estimate of the tidal disruption radius of the star can be determined by equating the \emph{maximum} self-gravitational field within the star to the tidal force of the SMBH (note that this differs from the usual definition, which equates the stellar surface gravity to the tidal acceleration), from which it follows that the distance from the SMBH at which the star is completely destroyed is approximately\footnote{In particular, Equation \eqref{rtc} approximates the functional form of the gravitational field within the star to analytically derive the radius where the self-gravitational field is maximized. Correspondingly, Equation \eqref{rtc} slightly underpredicts the value of $r_{\rm t, c}$ that is obtained by numerically determining the radius at which the self-gravitational field is maximized within the star (see Figure 2 of \citealt{coughlin22a}), and hence the ratio of the stream density to the SMBH density is somewhat larger than what is predicted using this approximation. For example, using the precise values in Table 2 of \citet{coughlin22a} shows that $\rho_{\rm i}/\rho_{\bullet} \simeq 5.5$ for a 5/3 polytrope.}

\begin{equation}
r_{\rm t, c} = r_{\rm t}\left(\frac{\rho_{\rm i}}{4\rho_{\star}}\right)^{-1/3}. \label{rtc}
\end{equation}
Here $r_{\rm t} = R_{\star}\left(M_{\bullet}/M_{\star}\right)^{1/3}$ is the usual tidal radius, with $R_{\star}$ and $M_{\star}$ the stellar radius and mass, and $\rho_{\rm i}$ ($\rho_{\star}$) is the initial central (average) density. Using this as the distance within which the star comes to be destroyed, it follows that the ratio of the central density of the expanding stream to the SMBH density at $r_{\rm t, c}$ is

\begin{equation}
\frac{\rho_{\rm i}}{\rho_{\bullet}} = 4. \label{rhocrhobullet}
\end{equation}
Thus, even though the star is destroyed at the distance given by Equation \eqref{rtc}, the stream is still self-gravitating. The reason for this seemingly contradictory conclusion is that the self-gravitational field of the star is maximized off-center, and hence the black hole density does not need to exceed the central density of the star to successfully destroy it.

Equation \eqref{rhocrhobullet} therefore shows that the stream is initially self-gravitating. To determine if it remains so, note that if the stream is in approximate hydrostatic balance, then from the Poisson equation the stream pressure, $p$, cross-sectional radius, $H$, and stream density are related via

\begin{equation}
\frac{p}{\rho H^2} \simeq 4\pi G\rho. \label{poisson}
\end{equation}
If the stream is adiabatic with adiabatic index $\gamma$, then $p \propto \rho^{\gamma}$. It also follows that $\rho \propto H^{-2}L^{-1}$, where $L$ is the length of the stream. If ballistic motion is approximately upheld along the length of the stream, then from the radial momentum equation it follows that $L \propto r^{2}$ near the marginally bound radius \citep{coughlin16}, which is just the growing term in the homologous solution to Equation \eqref{zmom1} when the pressure and self-gravity terms are neglected. There is also a branch that scales as $\propto r^{-1/2}$ that is important when the initial encounter of the star is very deep, as this sets the length of the stream at the time the star reaches pericenter (\citealt{stone13, darbha19}). For the outgoing evolution, initial conditions determine the relative contribution of each branch, but the solution that scales as $\propto r^2$ quickly dominates and the decaying solution is irrelevant (see also the discussion in \citealt{bonnerot22}). Combining these results with Equation \eqref{poisson}, it follows that the density along the axis of the stream and the stream cross-sectional radius scale with distance from the SMBH as

\begin{equation}
\rho \propto r^{-\frac{2}{\gamma-1}}, \quad H \propto r^{\frac{2-\gamma}{\gamma-1}}. \label{rhoceq}
\end{equation}
See also Equation (62) of \citet{coughlin16}.

From Equation \eqref{rhoceq}, the ratio of the stream density to the SMBH density scales as

\begin{equation}
\frac{\rho}{\rho_{\bullet}} \propto r^{\frac{3\gamma - 5}{\gamma-1}}.
\end{equation}
From this expression we see that $\gamma = 5/3$ demarcates the critical adiabatic index that allows the stream to remain self-gravitating as it recedes from the SMBH: for equations of state that satisfy $\gamma > 5/3$, the stream density will asymptotically outweigh the black hole density, and the latter is effectively ignorable as concerns the effects of self-gravity across the stream. For $\gamma < 5/3$, the stream density declines more rapidly than that of the SMBH, and eventually the stream will enter into a shear-dominated phase that yields $\sim$ homologous expansion (i.e., ballistic motion of the gas). 

A gas pressure-dominated stream with $\gamma \simeq 5/3$ is therefore interesting from a gravitational stability standpoint, as in this case the stream is just able to maintain quasi-hydrostatic balance. By quasi-hydrostatic we mean that the assumption of hydrostatic balance and the Poisson equation both being satisfied is consistent, even though the stream width expands with time and the background density declines with time (as $\rho \propto t^{-2}$). Of course, a $\gamma = 5/3$ equation of state is also quite relevant from a physical standpoint, as low-mass stars are overwhelmingly dominated by gas pressure (compared to radiation pressure, which would soften the equation of state). We would expect the $\gamma = 5/3$, adiabatic assumption to hold until the gas begins to recombine, which we discuss further in Section \ref{sec:recombination} below.

Therefore, the specific case of a $\gamma = 5/3$, cylindrical stream warrants further consideration and analysis, as concerns both the quasi-hydrostatic solution itself and its stability. In the next section we briefly justify the use of the coordinate system that we use throughout the remainder of the paper before continuing with this analysis.

\section{Coordinates and Equations}
\label{sec:coordinates}
\begin{figure}
   \centering
   \includegraphics[width=0.48\textwidth]{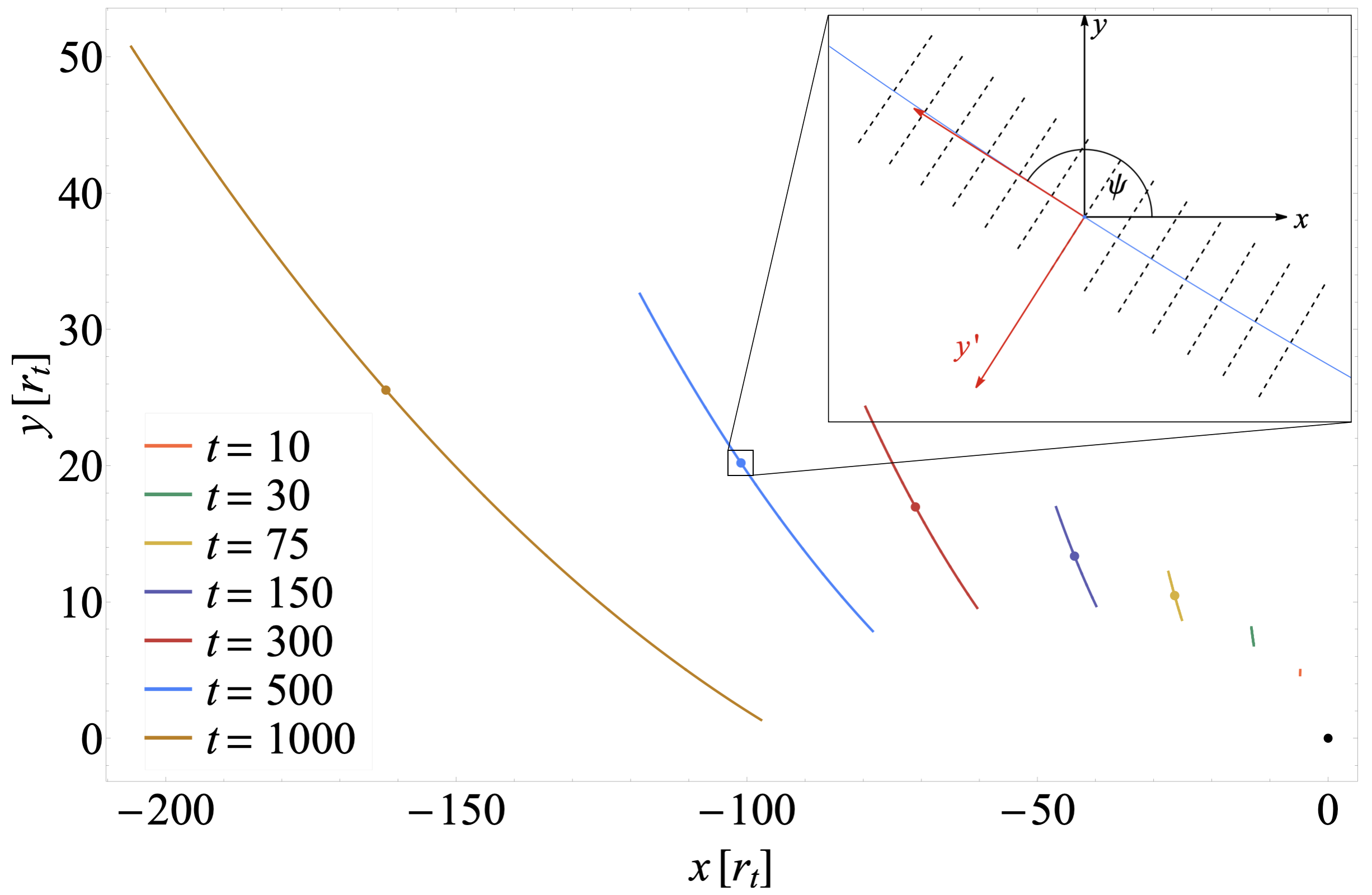} 
   \caption{The maximum-density ($X$-$Y$) curve from the frozen-in approximation for the times shown in the legend, where $t = 0$ corresponds to the time at which the star is at the tidal radius and times are measured in units of $r_{\rm t}^{3/2}/\sqrt{GM_{\bullet}}$, which is also the dynamical time of the original star. The points show the location of the marginally bound radius, and the inset gives an example of the coordinate system adapted to the stream. The dashed lines give the $y'$-direction at each point along the $X$-$Y$ curve, and $\psi$ is the angle between the $x$-axis and the direction along the curve (the red arrow at the point in the inset).}
   \label{fig:schema}
\end{figure}

At a given time $t$ there is a set of points along which the density of the stream material is maximized. Define this curve as $\{X,\,\,Y\}$ in the $x$-$y$ plane, where the $x$-axis points in the direction of pericenter of the disrupted star and $y$ is defined in a right-handed sense with respect to the angular momentum vector $\mathbf{\ell}$ of the original stellar orbit, i.e., $\mathbf{x}\times \mathbf{y} \propto \mathbf{\ell}$. We assume there is no torque out of the orbital plane\footnote{It is possible to include out-of-plane motion, which would be relevant if (e.g.) the black hole has spin and the spin direction is misaligned with respect to the orbital plane of the star. In this case the coordinate system is composed of the unit tangent vector, the unit signed curvature vector, and the unit binormal vector (in our simplified case the binormal is in the fixed direction of the angular momentum vector of the star).}. The $X-Y$ curve constitutes one curvilinear axis, and we define the $y'$-axis to be orthogonal to and within the plane of the $X-Y$ curve at any given point along that curve. Then the $x$-$y$ coordinates of any fluid element are related to $X$, $Y$, and $y'$ via
\begin{equation}
x = X-y'\sin\psi,
\quad y = Y+y'\cos\psi,
\end{equation}
where
\begin{equation}
\tan\psi = \frac{\partial Y}{\partial X}.
\end{equation}
The angle $\psi$ gives the direction tangent to the maximum-density curve relative to the fixed $x$-$y$ coordinate system, and thus defines ``along the stream,'' while $y'$ points ``transverse'' to the stream. Figure \ref{fig:schema} shows the relevant quantities calculated with the frozen-in approximation (see below for more details). 

The equations of motion within the orbital plane of the original star are straightforwardly derivable from the fluid equations by making this coordinate transformation, the result being
\begin{equation}
\cos\psi\frac{\partial^2Y}{\partial t^2}-\sin\psi\frac{\partial^2X}{\partial t^2}+\frac{\partial^2y'}{\partial t^2}-y'\left(\frac{\partial\psi}{\partial t}\right)^2+\frac{1}{\rho}\frac{\partial p}{\partial y'}+\frac{\partial \Phi}{\partial y'} = 0, \label{eqgen1}
\end{equation}
\begin{multline}
\cos\psi\frac{\partial^2X}{\partial t^2}+\sin\psi\frac{\partial^2Y}{\partial t^2}-2\frac{\partial y'}{\partial t}\frac{\partial \psi}{\partial t}-y'\frac{\partial^2\psi}{\partial t^2} \\ 
+\frac{\cos\psi}{1-y'\cos\psi\frac{\partial \psi}{\partial X}}\left(\frac{1}{\rho}\frac{\partial p}{\partial X}+\frac{\partial \Phi}{\partial X}\right) = 0, \label{eqgen2}
\end{multline}
The gravitational potential $\Phi$ has contributions from self-gravity and from the black hole; the latter is
\begin{equation}
\Phi_{\bullet} = -\frac{GM_{\bullet}}{\sqrt{X^2+Y^2+\left(y'\right)^2+2y'\left(Y\cos\psi-X\sin\psi\right)}}.
\end{equation}
At early times (which we quantify below) $y'\ll \sqrt{X^2+Y^2}$, and we Taylor expand the gravitational potential in powers of $y'/\sqrt{X^2+Y^2}$ to second order, take the derivative, and maintain terms up to first order (the tidal approximation). Doing so gives
\begin{multline}
\frac{\partial\Phi_{\bullet}}{\partial y'} = \frac{GM_{\bullet}}{\left(X^2+Y^2\right)^{3/2}}\bigg\{-X\sin\psi+Y\cos\psi \\ 
+\left(1-3\frac{\left(X\sin\psi-Y\cos\psi\right)^2}{X^2+Y^2}\right)y'\bigg\}
\end{multline}
\begin{multline}
\frac{\partial\Phi_{\bullet}}{\partial X} = \frac{GM_{\bullet}}{\left(X^2+Y^2\right)^{3/2}}\bigg\{X+Y\tan\psi \\ 
-y'\left(3\frac{\left(Y\cos\psi-X\sin\psi\right)\left(X+Y\tan\psi\right)}{X^2+Y^2}+\left(Y\sin\psi+X\cos\psi\right)\frac{\partial \psi}{\partial X}\right)\bigg\}
\end{multline}

Inserting these expressions into Equations \eqref{eqgen1} and \eqref{eqgen2} gives the general equations of motion for fluid elements within the stream and within the orbital plane of the original star, which contain Coriolis and centrifugal terms that depend on temporal derivatives of $\psi$, and also curvature terms that are proportional to $\partial \psi/\partial X$. The magnitude of each of these terms is related to the angular momentum of the fluid, for if the angular momentum were precisely zero, we could set $Y \equiv 0$ ($\psi \equiv 0$) without loss of generality. To gain an understanding of the importance of these non-inertial and curvilinear terms, we can assume that the maximum-density curve solves the dynamical equations of motion, i.e., $X$ and $Y$ satisfy
\begin{equation}
\frac{\partial^2X}{\partial t^2} = -\frac{GM_{\bullet}X}{\left(X^2+Y^2\right)^{3/2}}, \quad \frac{\partial^2Y}{\partial t^2} = -\frac{GM_{\bullet}Y}{\left(X^2+Y^2\right)^{3/2}}. \label{XcYc}
\end{equation}
Note that these follow directly from Equations \eqref{eqgen1} and \eqref{eqgen2} when $y' \equiv 0$ if we ignore pressure and self-gravity, meaning that they should be approximately upheld in general while the stream is thin. If we make the frozen-in approximation \citep{lacy82} with the pericenter distance of the original star equal to the tidal radius $r_{\rm t}$, where $r_{\rm t} = R_{\star}\left(M_{\bullet}/M_{\star}\right)^{1/3}$ with $R_{\star}$ and $M_{\star}$ the stellar radius and mass, then the initial conditions are $X(X_0, t = 0) = X_0$, $\dot{X}(t = 0) = 0$, $Y(t = 0) = 0$, and $\dot{Y}(t = 0) = \sqrt{2GM/r_{\rm t}}$ with $r_{\rm t}-R_{\star} \le X_0 \le r_{\rm t}+R_{\star}$ (dots denote temporal derivatives). The initial position of a fluid element is also proportional to its specific energy, with $X_0 < r_{\rm t}$ being bound, $X_0 > r_{\rm t}$ unbound, and $X_0 = r_{\rm t}$ marginally bound (i.e., on a parabolic orbit). We can then integrate the equations of motion and directly assess the magnitude of the non-inertial terms.

\begin{figure} 
   \centering
   \includegraphics[width=0.47\textwidth]{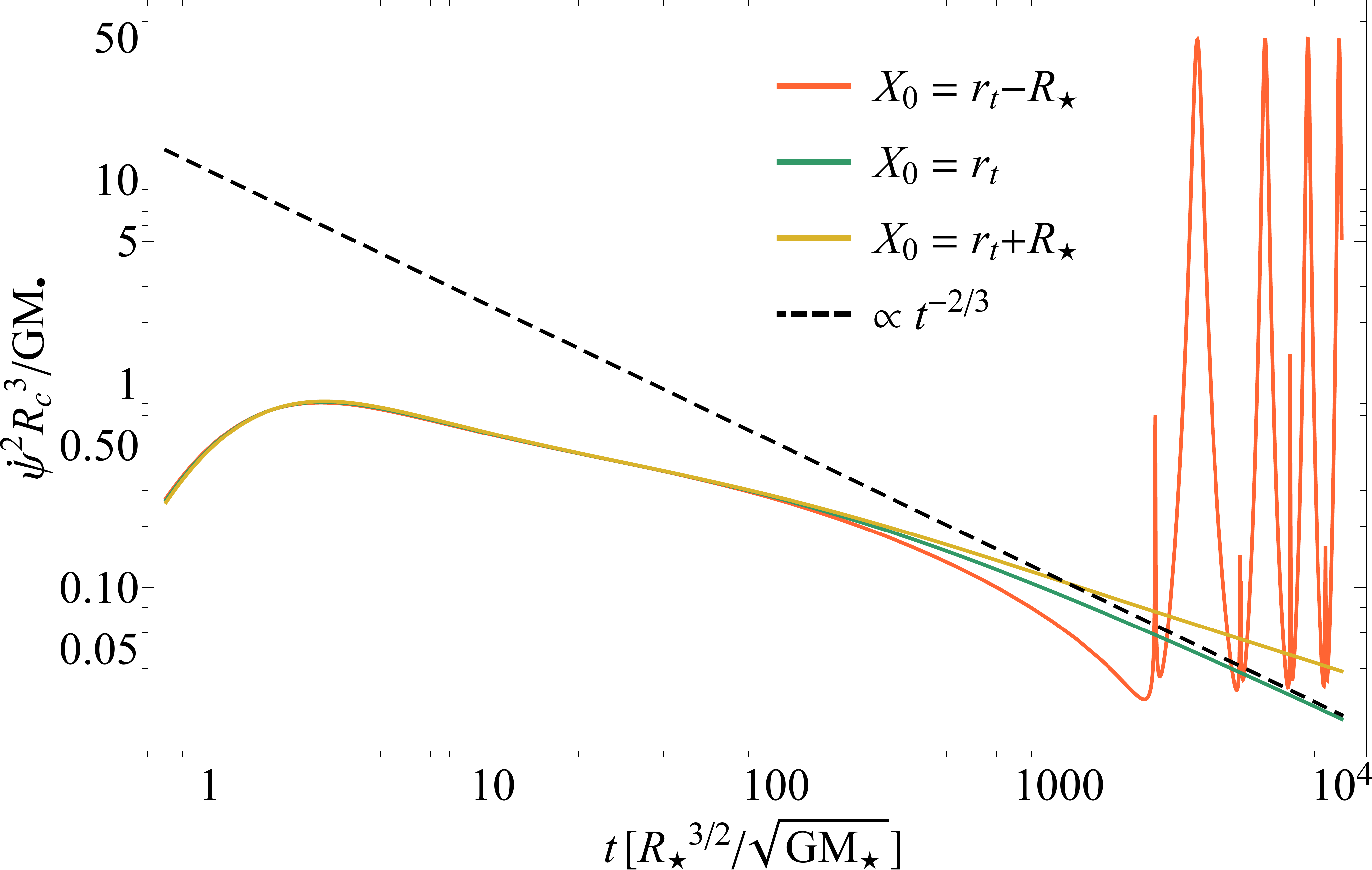} 
   \caption{The ratio of the centrifugal term to the tidal force as a function of time for three different initial positions within the star, which correspond to the most-bound fluid element ($X_0 = r_{\rm t}-R_{\star}$, the marginally bound fluid element ($X_0 = r_{\rm t}$), and the most-unbound fluid element ($X_0 = r_{\rm t}+R_{\star}$).}
   \label{fig:centrifugal}
\end{figure}

Figure \ref{fig:schema} shows the $\{X,\,Y\}$ curve calculated with this approach at the times in the legend, where $t = 0$ corresponds to when the star is at $r_{\rm t}$ and times are in units of $r_{\rm t}^{3/2}/\sqrt{GM_{\bullet}} = R_{\star}^{3/2}/\sqrt{GM_{\star}}$, which is the dynamical time of the original star. Here we let $M_{\bullet}/M_{\star} = 10^6$. The inset gives an example of the $y'$ coordinate system at $t = 500$, and the points show the location of the zero-energy orbit for which $X_0 = r_{\rm t}$. 

The non-inertial term on the left-hand side of Equation \eqref{eqgen1} that modifies the $y'$ equation of motion relative to the tidal term is $\dot{\psi}^2R^{3}/(GM_{\bullet})$, where $R = \sqrt{X^2+Y^2}$. Figure \ref{fig:centrifugal} shows this ratio as a function of time for three different initial positions along the maximum-density curve: $r_{\rm t}-R_{\star}$ being the most-bound fluid element, $r_{\rm t}$ being marginally bound, and $r_{\rm t}+R_{\star}$ being the most unbound. All three curves are nearly indistinguishable until $t \simeq 2000$, at which time the most-bound fluid element returns to the SMBH and the centrifugal terms are important (this also violates the tidal approximation, as by this time the length of the stream is comparable to $\sqrt{X^2+Y^2}$). At sufficiently late times the marginally bound solution decays approximately as $\propto t^{-2/3}$, which is shown by the black, dashed line.

\begin{figure} 
   \centering
   \includegraphics[width=0.47\textwidth]{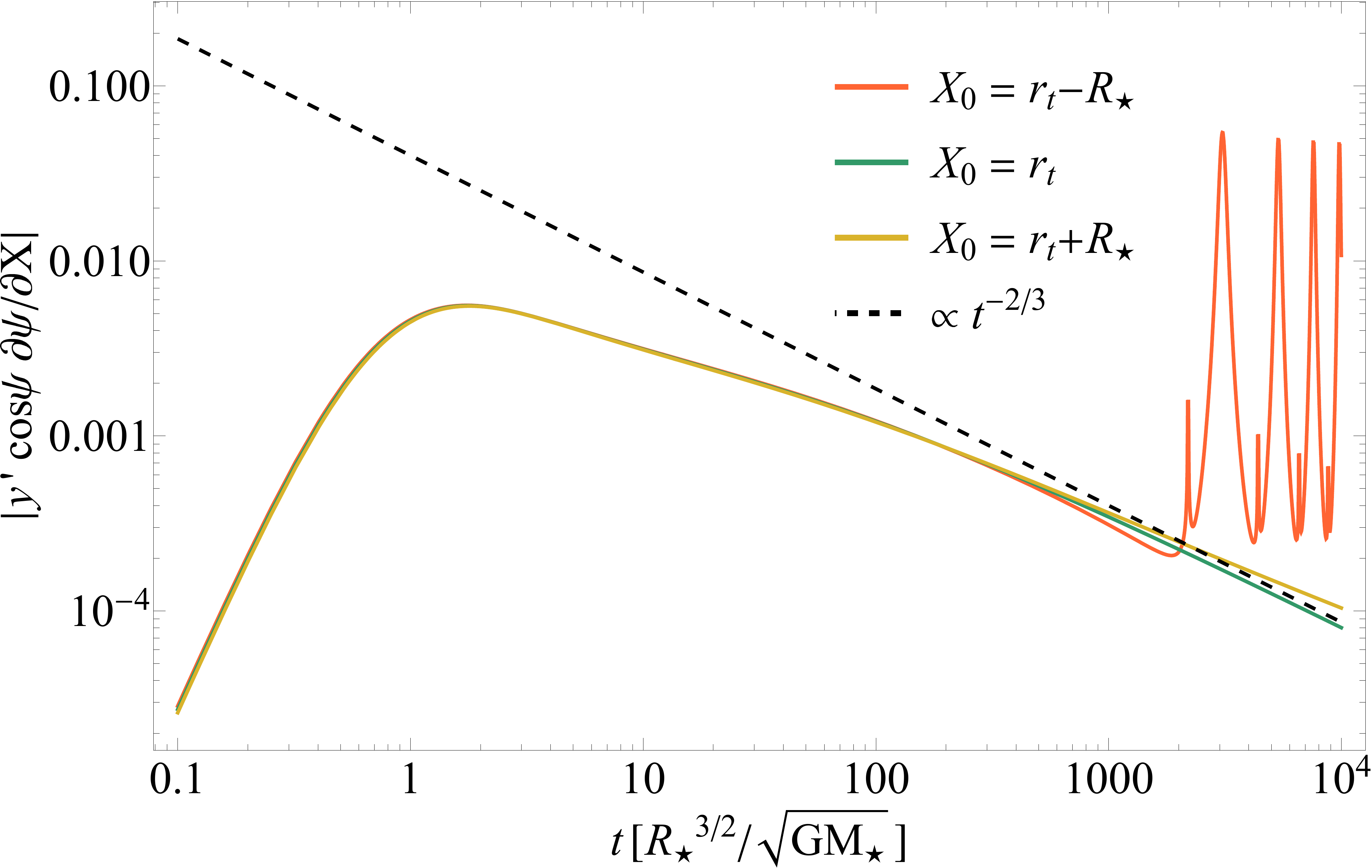} 
   \caption{The curvature term in the equations of motion under the frozen-in approximation, where here $M_{\bullet} = 10^6M_{\odot}$ and $R_{\star} = R_{\odot}$. Here we let $y' = R_{\star}\left(R/r_{\rm t}\right)^{1/2}$, which is the expected scaling for the width of the stream, and $X$ is measured in units of $r_{\rm t}$. At all times this quantity is small and decays with time for the unbound segment of the stream, and only increases in importance when the stream material returns to pericenter (for the most-bound fluid element this occurs around $\sim 2000$ days post-disruption). }
   \label{fig:curvature}
\end{figure}

The curvature term that modifies the spatial derivatives (the left-hand side of Equation \ref{eqgen2}) and that contributes a correction to the tidal term (the right-hand side of Equation \ref{eqgen2}) is $y'\cos\psi \partial\psi/\partial X$. This term represents the fact that the distance between adjacent $y'$-axes will change along the $\{X,\,Y\}$ curve if $\partial\psi/\partial X \neq 0$. From the discussion in Section \ref{sec:basic} and as we show more rigorously in the next section, the transverse extent of the stream is much less than the distance to any given fluid element, and from Equation \eqref{rhoceq} we expect fluid elements to approximately satisfy $y' \simeq R_{\star}\left(R/r_{\rm t}\right)^{1/2}$ for a gas-pressure dominated equation of state. Figure \ref{fig:curvature} shows $|y'\cos\psi \partial \psi/\partial X|$ with $y' = R_{\star}\left(R/r_{\rm t}\right)^{1/2}$ as a function of time for the same three initial positions and the same frozen-in approximation as used in Figure \ref{fig:centrifugal}, and $X$ is measured in units of $r_{\rm t}$. We see that this term is always smaller than one, and decays with time for the marginally bound and unbound segments of the stream, the former scaling as $\propto t^{-2/3}$ at late times. The curvature increases in importance once the most-bound segment of the stream returns to pericenter, around $\sim 2000$ days post-disruption.

Figures \ref{fig:centrifugal} and \ref{fig:curvature} show that the non-inertial terms increase in importance for $\lesssim 2-3$ dynamical times and thereafter decay, and although the frozen-in approximation does not yield the correct dynamical evolution of the stream in detail \citep{steinberg19}, this suggests that they are relevant for only a few dynamical times post-disruption. For $\lesssim 1000$ dynamical times the entire stream behaves effectively as if it is marginally bound and the non-inertial terms (relative to the tidal terms) decay with time, and we can always find a region nearer the marginally bound radius where the non-inertial terms are ignorable to yet-later times. 

For the remainder of the paper we focus on this region of parameter space, and we approximate $Y \simeq 0$ ($\psi \simeq 0$). We also define $X = R_{\rm c}(t)+z(z_0,t)$ with $R_{\rm c}$ the zero-energy Keplerian orbit that satisfies

\begin{equation}
\frac{\partial R_{\rm c}}{\partial t} = \sqrt{\frac{2GM_{\bullet}}{R_{\rm c}}} \quad \Leftrightarrow \quad R_{\rm c}(t) = R_{\rm i}\left(1+\frac{3}{2}\frac{\sqrt{2GM_{\bullet}}}{R_0^{3/2}}t\right)^{2/3}, \label{Rcoft}
\end{equation}
where $R_{\rm i}$ is an arbitrary scale radius. The out-of-plane equation of motion is the same as that for the $y'$-direction; if we change to cylindrical coordinates and also ignore angular variations around the stream (generalizing the solutions to include these perturbations is straightforward), then the change to cylindrical coordinates amounts to letting $y'\rightarrow s$ with $s$ the cylindrical radius. The dynamical equations are therefore
\begin{equation}
\frac{\partial^2z}{\partial t^2}+\frac{1}{\rho}\frac{\partial p}{\partial z}+\frac{\partial \Phi}{\partial z} = \frac{2GM_{\bullet}}{R_{\rm c}^3}z, \label{zmom1}
\end{equation} 
\begin{equation}
\frac{\partial^2s}{\partial t^2}+\frac{1}{\rho}\frac{\partial p}{\partial s}+\frac{\partial \Phi}{\partial s} = -\frac{GM_{\bullet}}{R_{\rm c}^3}s. \label{smom1}
\end{equation}
These are what we would have postulated immediately given the discussion and motivation presented in the previous section, but it is useful to have this more general definition when making comparisons to simulations where the coordinates of the maximum-density curve are more complex. We also need the Poisson equation for the self-gravitational field, which is
\begin{equation}
\frac{1}{s}\frac{\partial}{\partial s}\left[s\frac{\partial \Phi}{\partial s}\right]+\frac{\partial^2\Phi}{\partial z^2} = 4\pi G\rho. \label{phi1}
\end{equation}
The conservation of mass in Lagrangian form is
\begin{equation}
\rho J \frac{s}{s_0} = \rho_{\rm i}g_0(s_0,z_0), \label{rhoLag}
\end{equation}
where $J = |\partial x^{i}/\partial x_0^{j}|$ is the Jacobian that transforms between the current $\{s,z\}$ and initial $\{s_0,z_0\}$ Lagrangian positions, $\rho_{\rm i}$ is a scale density, and $g_0(s_0,z_0)$ is the dimensionless initial density profile. Since we are ignoring angular variations the Jacobian is
\begin{equation}
J = \frac{\partial s}{\partial s_0}\frac{\partial z}{\partial z_0}-\frac{\partial s}{\partial z_0}\frac{\partial z}{\partial s_0}. \label{jac}
\end{equation}
We also assume for now that the fluid is adiabatic; in this case the entropy is a conserved Lagrangian quantity, and hence
\begin{equation}
p = p_{\rm i}K_{0}(s_0,z_0)\left(\frac{\rho}{\rho_{\rm i}}\right)^{\gamma}, \label{eosad}
\end{equation}
where $p_{\rm i}$ is a scale pressure, $K_{0}(s_0,z_0)$ is the dimensionless entropy function of the gas, and $\gamma$ is the adiabatic index of the gas. 

In the next section we show that there is an exact solution (which accounts for the transverse structure of the debris stream) to Equations \eqref{zmom1} -- \eqref{eosad} when the stream is cylindrically symmetric and the fluid is gas-pressure dominated with $\gamma = 5/3$.

\section{Exact Solutions}
\label{sec:exact}
We expect solutions to Equations \eqref{zmom1} -- \eqref{rhoLag} to be approximately cylindrically symmetric with $\partial/\partial z \simeq 0$. From the discussion in Section \ref{sec:basic}, the cross-sectional radius of the stream should also expand roughly as $\propto R_{\rm c}^{1/2}$, which results from the confinement by self-gravity coupled to the declining stream density. By inspection we see that the following has these properties and exactly solves Equations \eqref{zmom1} -- \eqref{rhoLag} when $\gamma = 5/3$ and the density, pressure, and self-gravitational potential are cylindrically symmetric:
\begin{equation}
\begin{split}
z &= H_{\rm i}z_0e^{2\tau}, \,\,\, s = H_{\rm i}e^{\tau/2}s_0, \,\,\, \rho = \rho_{\rm i} e^{-3\tau} g_0(s_0), \\ 
p &= p_{\rm i}K_0(s_0)e^{-5\tau}g_0(s_0)^{5/3}, \,\,\, \Phi = \frac{p_{\rm i}}{\rho_{\rm i}}e^{-2\tau}j_0(s_0). \label{exactsol}
\end{split}
\end{equation}
Here $s_0$ and $z_0$ are measured relative to the initial cylindrical radius of the stream $H_{\rm i}$, $\rho_{\rm i}$ and $p_{\rm i}$ are the density and pressure along the stream axis at the time when $R_{\rm c}(t) = R_{\rm i}$, and
\begin{equation}
\tau = \ln\left(\frac{R_{\rm c}(t)}{R_{\rm i}}\right).
\end{equation}
The dimensionless density $g_0$, dimensionless entropy $K_0$, and dimensionless gravitational potential $j_0$ are related by the equation of hydrostatic equilibrium and the Poisson equation, which are respectively
\begin{equation}
\frac{1}{g_0}\frac{d}{d s_0}\left[K_0 g_0^{5/3}\right] = -\frac{d j_0}{d s_0}, \label{hse}
\end{equation} 
\begin{equation}
\frac{p_{\rm i}}{\rho_{\rm i}H_{\rm i}^2}\frac{1}{s_0}\frac{d}{ds_0}\left[s_0\frac{d j_0}{ds_0}\right] = 4\pi G\rho_{\rm i}g_0, \label{poissonaux}
\end{equation}
and can be combined to give
\begin{equation}
\alpha_{\rm i}\frac{1}{s_0}\frac{d}{ds_0}\left[s_0\frac{1}{g_0}\frac{d}{ds_0}\left[K_0(s_0)g_0^{5/3}\right]\right] = -g_0, \label{HSE2}
\end{equation}
where
\begin{equation}
\alpha_{\rm i} \equiv \frac{p_{\rm i}}{\rho_{\rm i}H_{\rm i}^2}\frac{1}{4\pi G\rho_{\rm i}}. \label{Kstardef}
\end{equation}

\begin{figure}
   \centering
   \includegraphics[width=0.475\textwidth]{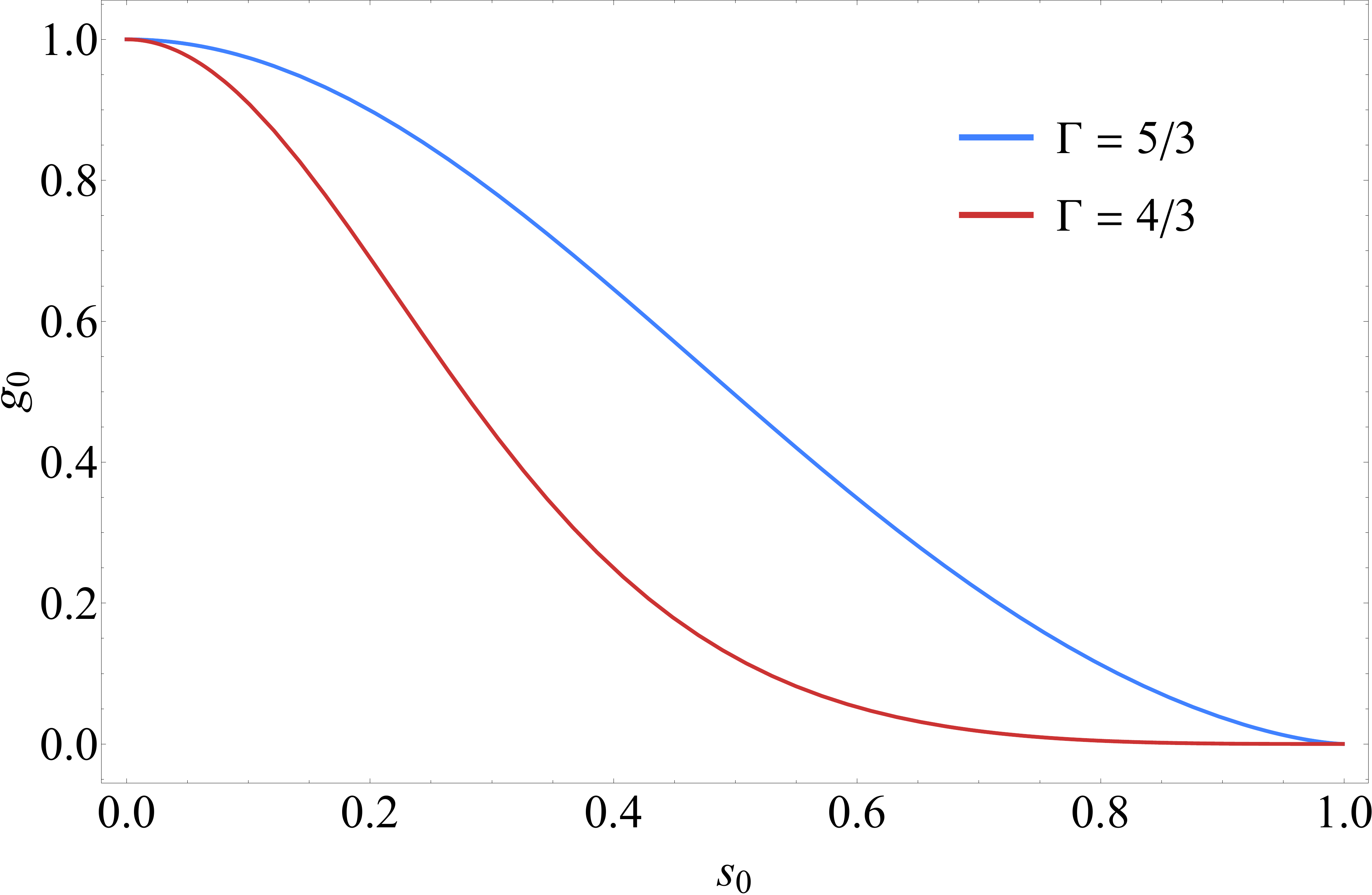} 
   \caption{The dimensionless density profile that satisfies the cylindrical Lane-Emden equation, i.e., a polytropic density profile in cylindrical coordinates, with a polytropic index of 5/3 (blue) and 4/3 (red).}
   \label{fig:hse}
\end{figure}

Equation \eqref{Kstardef} cannot be solved without an additional prescription for the entropy profile. There are two cases that are (especially) relevant to TDEs, the first of which is for low-mass stars that are fully convective and that therefore have effectively constant entropy; in this case $K_0(s_0) \equiv 1$. The other case is for higher-mass and radiative stars, which -- at least for stars near the zero-age main sequence and that are not highly evolved -- can be accurately modeled by the Eddington standard model. In this case, the entropy profile as a function of {spherical radius} of the initial star is $K_0 = g_0^{-1/3}$ (the Eddington standard model has a pressure profile $p \propto \rho^{4/3}$, and from Equation \eqref{eosad} the entropy therefore scales as $K_0 \propto \rho^{-1/3}$). Clearly after the star is tidally destroyed the entropy profile will exhibit variation both with cylindrical radius and along the stream. However, the shear along the stream axis is much greater than that perpendicular to it, and hence variations of the entropy profile with $z_0$ should be much smaller than those in $s_0$ near the marginally bound radius. Furthermore, since the pressure is maximized at the center of the original star, the entropy profile must satisfy $\partial K_0/\partial z_0 (z_0 = 0) = 0$. Thus, within the set of approximations we have already made, the relation $K_0 = g_0^{-1/3}$ should also hold for the disruption of a radiative star, where $g_0$ is a function of cylindrical radius (and for any arbitrary initial entropy profile this approximation should be upheld to a good degree of accuracy). We therefore have 
\begin{equation}
\alpha_{\rm i}\frac{1}{s_0}\frac{d}{ds_0}\left[\frac{s_0}{g_0}\frac{d}{ds_0}\left[g_0^{\Gamma}\right]\right] = -g_0, \label{lanecyl}
\end{equation}
where $\Gamma = 5/3$ for a low-mass (convective) star and $\Gamma = 4/3$ for a high-mass (radiative) star. Equation \eqref{lanecyl} is just the cylindrical Lane-Emden equation (e.g., \citealt{ostriker64}), and the solution to it must satisfy $g_0(0) = 1$ and $dg/ds_0(0) = 0$. Since the surface at which $g_0 = 0$ coincides with $s_0 = 1$ by definition, there is a value of $\alpha_{\rm i}$ that will simultaneously satisfy all three of these boundary conditions; numerically we find
\begin{equation}
\alpha_{\rm i}(\Gamma = 5/3) \simeq 0.0571, \,\,\, \alpha_{\rm i}(\Gamma = 4/3) \simeq 0.0196. \label{Kstar}
\end{equation}
Figure \ref{fig:hse} shows the numerical solution to Equation \eqref{lanecyl} for $\Gamma = 5/3$ (blue) and $\Gamma = 4/3$ (red). Analogously to spherical polytropes, the more compressible solution with $\Gamma = 4/3$ has a more rarefied envelope compared to the solution with $\Gamma = 5/3$. 

The solution given by Equation \eqref{exactsol} possesses homologous expansion in both the cylindrical-radial direction and along the axis of the filament, but at different rates. Notice that these solutions show that, consistent with Equation \eqref{rhoceq} above, the radius of the stream expands as
\begin{equation}
H = H_{\rm i}\left(\frac{R_{\rm c}}{R_{\rm i}}\right)^{1/2} \propto t^{1/3}.
\end{equation} 
The solution \eqref{exactsol} represents the quasi-hydrostatic ``background,'' or ``equilibrium'' solution, where the dynamical expansion of the stream is consistent with both the tidal force of the SMBH and the confinement by self-gravity. The fact that such a solution exists is, in essence, a more direct and rigorous demonstration of the validity of the more heuristic arguments in Section \ref{sec:basic}, where it was shown that the stream density and SMBH ``density'' scale identically when $\gamma = 5/3$ and the stream is also self-gravitating. This equilibrium will only be established for a very specific set of initial conditions, in particular those that have the initial pressure, density, and stream width related by Equation \eqref{Kstardef}. In the next section we analyze perturbations on top of this background state to understand its stability in the presence of more general initial conditions. 

\section{Cylindrically symmetric perturbations and critical stream density}
\label{sec:perturbations}
We treat the previously derived solution as the equilibrium state that we perturb. Here we focus only on the case where the perturbations are cylindrically symmetric and do not possess variation along the axis of the stream; as we now demonstrate, cylindrically symmetric perturbations alone yield a fundamental stability criterion as concerns the ability of the stream to remain self-gravitating. 

\begin{figure*} 
   \centering
   \includegraphics[width=0.495\textwidth]{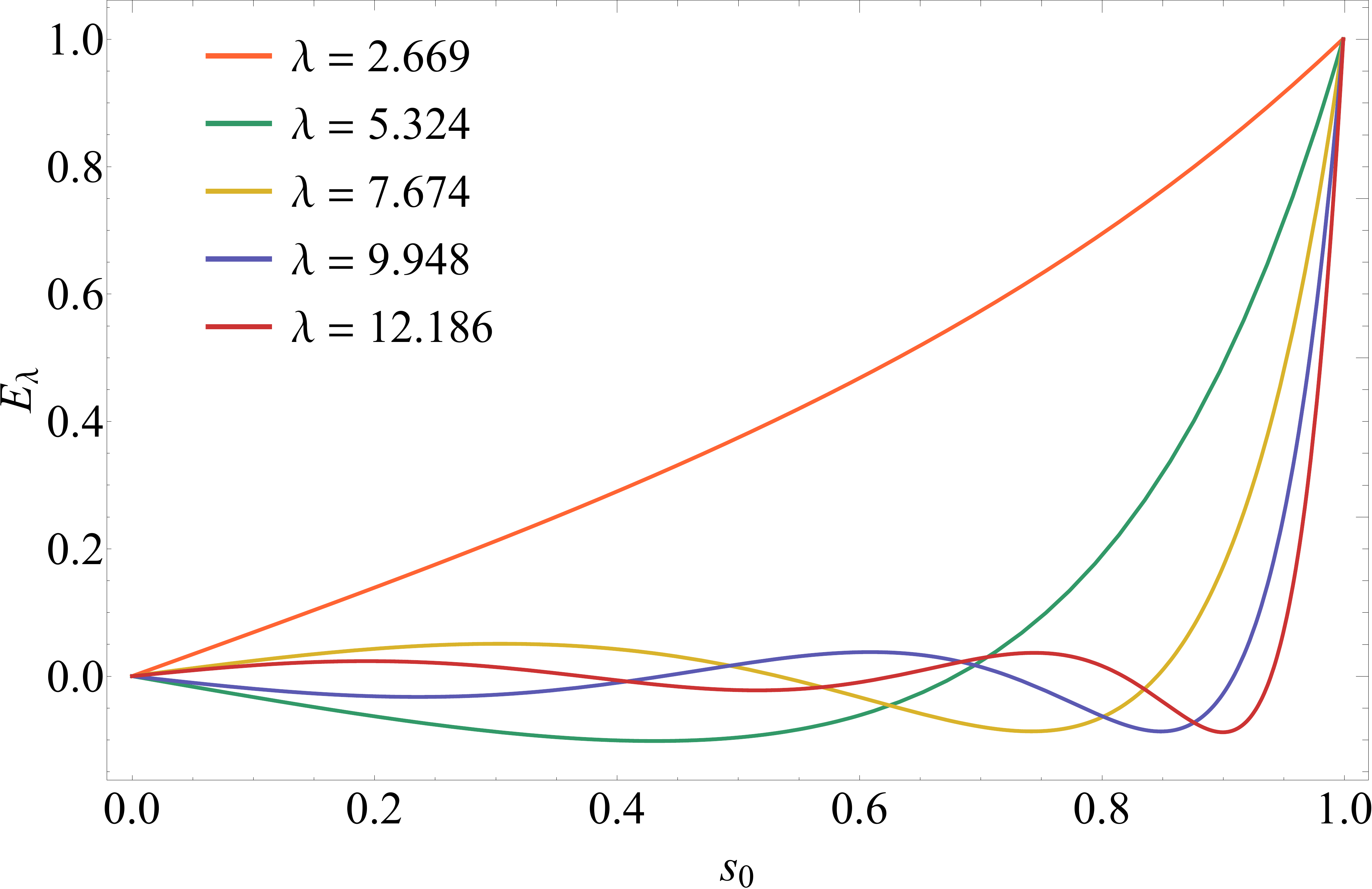} 
   \includegraphics[width=0.495\textwidth]{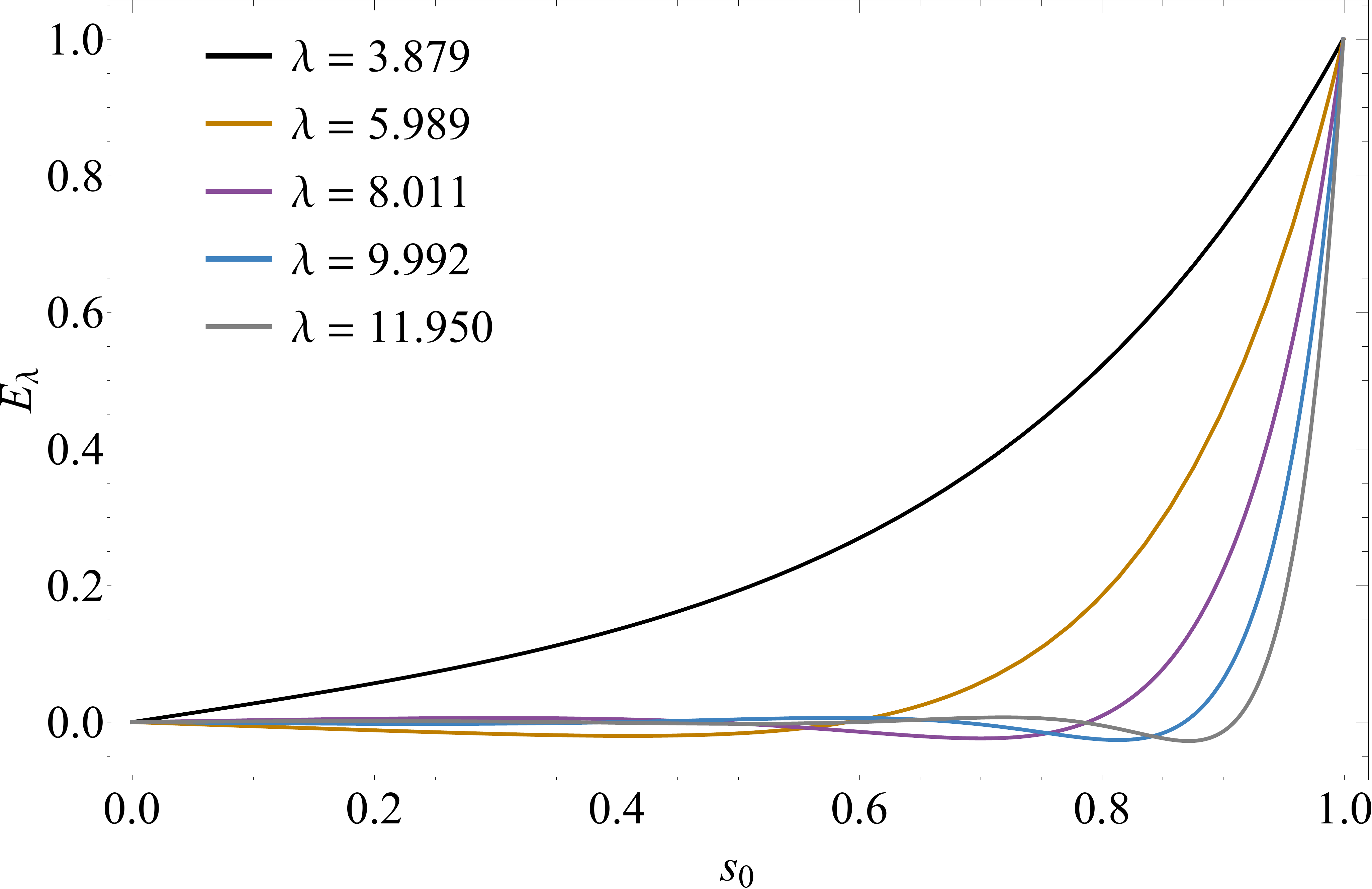} 
   \caption{The first five eigenmodes of a cylinder with polytropic index $\Gamma = 5/3$ (left) and $\Gamma = 4/3$ (right). The eigenvalues are shown in the legend. The lowest-order mode is the analog of the ``breathing mode'' exhibited by stars undergoing spherically symmetric oscillations, such that the motion is purely inward or outward as a function of time, while the higher-order modes are ``overtones'' that have both inward and outward motion at a single time.}
   \label{fig:eigenfuns}
\end{figure*}

To account for cylindrically symmetric perturbations we decompose the Lagrangian positions of fluid elements as
\begin{equation}
s = H_{\rm i}e^{\tau/2}\left\{s_0+s_1(s_0,\tau)\right\}, \label{spert}
\end{equation} 
where $s_1$ is an assumed-small correction to the initial Lagrangian position. Since we are restricting our analysis to cylindrically symmetric perturbations for which $\partial s/\partial z_0 = 0$, we have (note that the factors of $H_{\rm i}$ do not appear here because the Jacobian is dimensionless)
\begin{equation}
\begin{split}
J\frac{s}{s_0} = \frac{\partial s}{\partial s_0}\frac{\partial z}{\partial z_0}\frac{s}{s_0} &= e^{3\tau}\left(1+\frac{\partial s_1}{\partial s_0}\right)\left(1+\frac{s_1}{s_0}\right) \\ 
&=e^{3\tau}\left(1+\frac{1}{s_0}\frac{\partial}{\partial s_0}\left[s_0s_1\right]\right),
\end{split}
\end{equation}
where the last equality is correct to first order. The density is then, from Equation \eqref{rhoLag},
\begin{equation}
\begin{split}
\rho &= \rho_{\rm i}e^{-3\tau}g_0(s_0)\left(1+\frac{1}{s_0}\frac{\partial}{\partial s_0}\left[s_0 s_1\right]\right)^{-1} \\
&=  \rho_{\rm i}e^{-3\tau}g_0(s_0)\left(1-\frac{1}{s_0}\frac{\partial}{\partial s_0}\left[s_0 s_1\right]\right), \label{rhopert}
\end{split}
\end{equation}
and the final equality results from a first-order series expansion in subscript-1 quantities. We also perturb the self-gravitational potential by writing
\begin{equation}
\Phi = \frac{p_{\rm i}}{\rho_{\rm i}}e^{-2\tau}\left\{j_0(s_0)+j_1(s_0,\tau)\right\}. \label{Phidef}
\end{equation}
From the entropy equation \eqref{eosad}, the pressure to first order is
\begin{equation}
\begin{split}
p &= p_{\rm i}K_0(s_0)\left(\frac{\rho}{\rho_{\rm i}}\right)^{5/3} = p_{\rm i}e^{-5\tau} K_0(s_0)g_0^{5/3}\left(1-\frac{1}{s_0}\frac{\partial}{\partial s_0}\left[s_0s_1\right]\right)^{5/3} \\
 &= p_{\rm i}e^{-5\tau}K_0(s_0)g_0^{5/3}\left(1-\frac{5}{3}\frac{1}{s_0}\frac{\partial}{\partial s_0}\left[s_0s_1\right]\right). 
\end{split}
\end{equation}
We can now insert these definitions and change of variables into Equations \eqref{smom1} and \eqref{phi1} (note that Equation \ref{zmom1} is trivially satisfied by the background solution with no perturbations along the stream axis) and keep only leading-order terms. In Equation \eqref{smom1}, the dynamical and tidal terms combine to yield the following first-order correction:
\begin{equation}
\frac{\partial^2s}{\partial t^2}+\frac{2GM_{\bullet}}{R_{\rm c}^3}s = \frac{GM_{\bullet}H_{\rm i}e^{\tau/2}}{R_{\rm c}^3}\left(2\frac{\partial^2s_1}{\partial \tau^2}-\frac{\partial s_1}{\partial \tau}\right). \label{auxeq1}
\end{equation}
In the same equation the pressure gradient term is given by
\begin{equation}
\frac{1}{\rho}\frac{\partial p}{\partial s} = \frac{5}{2}\frac{p_{\rm i}}{\rho_{\rm i}H_{\rm i}}e^{-5\tau/2}K_0^{3/5}\frac{\partial}{\partial s_0}\left[K_0^{2/5}\left(\frac{\rho}{\rho_{\rm i}}\right)^{2/3}\right],
\end{equation}
and combining this with the gradient of the gravitational potential and using the equation of hydrostatic balance \eqref{hse} to cancel the zeroth-order terms yields
\begin{multline}
\frac{1}{\rho}\frac{\partial p}{\partial s}+\frac{\partial \Phi}{\partial s} \\ 
= \frac{p_{\rm i}}{\rho_{\rm i}H_{\rm i}}e^{-5\tau/2}\left\{\frac{\partial j_1}{\partial s_0}-\frac{5}{3}K_0^{3/5}\frac{\partial}{\partial s_0}\left[K_0^{2/5}\frac{g_0^{2/3}}{s_0}\frac{\partial}{\partial s_0}\left[s_0s_1\right]\right]\right\}. \label{auxeq2}
\end{multline}
Adding Equations \eqref{auxeq1} and \eqref{auxeq2} and setting the result to zero yields the first-order momentum equation:
\begin{multline}
2\frac{\partial^2s_1}{\partial \tau^2}-\frac{\partial s_1}{\partial \tau} \\ 
+\mu^2\left\{\frac{\partial j_1}{\partial s_0}-\frac{5}{3}K_0^{3/5}\frac{\partial}{\partial s_0}\left[K_0^{2/5}\frac{g_0^{2/3}}{s_0}\frac{\partial}{\partial s_0}\left[s_0s_1\right]\right]\right\} = 0, \label{auxeq3}
\end{multline}
where
\begin{equation}
\mu^2 \equiv \frac{R_{\rm i}^3}{GM_{\bullet}H_{\rm i}^2}\frac{p_{\rm i}}{\rho_{\rm i}} = 
\frac{3\rho_{\rm i}}{\rho_{\bullet}}\alpha_{\rm i}. \label{mudef}
\end{equation}
In the final equality in Equation \eqref{mudef} we used Equation \eqref{Kstar} and defined the ``black hole density'' by $\rho_{\bullet} = 3M_{\bullet}/(4\pi R_{\rm i}^3)$. Note that this is the same black hole density that was defined in Section \ref{sec:basic}, but here we are evaluating it at the scale radius $R_{\rm i}$.

To linearize the Poisson equation \eqref{phi1}, note from Equation \eqref{spert} that
\begin{equation}
\frac{\partial}{\partial s} = \frac{1}{H_{\rm i}e^{\tau/2}}\left(1-\frac{\partial s_1}{\partial s_0}\right)\frac{\partial}{\partial s_0}
\end{equation}
to first order, and hence -- for cylindrically symmetric perturbations -- the Poisson equation becomes, after using Equations \eqref{rhopert} and \eqref{Phidef} for the density and the gravitational potential,
\begin{equation}
\frac{p_{\rm i}}{\rho_{\rm i}H_{\rm i}^2}\frac{1}{s_0}\frac{\partial}{\partial s_0}\left[s_0\left(1-s_0\frac{\partial}{\partial s_0}\left[\frac{s_1}{s_0}\right]\right)\frac{\partial}{\partial s_0}\left[j_0+j_1\right]\right] = 4\pi G\rho_{\rm i}g_0(s_0).
\end{equation}
Using Equation \eqref{poissonaux} to eliminate the zeroth-order terms, this becomes (to first order)
\begin{equation}
\frac{\partial}{\partial s_0}\left[s_0\left(\frac{\partial j_1}{\partial s_0}-s_0\frac{\partial}{\partial s_0}\left[\frac{s_1}{s_0}\right]\frac{\partial j_0}{\partial s_0}\right)\right] = 0.
\end{equation}
Maintaining the regularity of the gravitational potential along the axis, we can integrate this equation to yield
\begin{equation}
\frac{\partial j_1}{\partial s_0} = s_0\frac{\partial}{\partial s_0}\left[\frac{s_1}{s_0}\right]\frac{\partial j_0}{\partial s_0}.
\end{equation}
We can now insert this result into Equation \eqref{auxeq3} and take the Laplace transform, where the Laplace transform of $s_1$ is
\begin{equation}
\tilde{s}_1(\sigma, s_0) = \int_0^{\infty}s_1(\tau,s_0)e^{-\sigma\tau}d\tau,
\end{equation}
If we let the initial velocity profile of the fluid be $\partial s_1/\partial \tau(\tau = 0) = V_0(s_0)$ (note that $\tilde{s}_1(\tau = 0) = 0$ by definition), then doing so yields the following equation for $\tilde{s}_1$:
\begin{equation}
\sigma\left(2\sigma-1\right)\tilde{s}_1-\mu^2\mathcal{L}\left[\tilde{s}_1\right] = 2V_0(s_0), \label{s1eq} 
\end{equation}
where
\begin{equation}
\mathcal{L}\left[\tilde{s}_1\right] = \frac{5}{3}K_0^{\frac{3}{5}}\frac{\partial}{\partial s_0}\left[K_0^{\frac{2}{5}}\frac{g_0^{2/3}}{s_0}\frac{\partial}{\partial s_0}\left[s_0\tilde{s}_1\right]\right]-s_0\frac{\partial}{\partial s_0}\left[\frac{\tilde{s}_1}{s_0}\right]\frac{\partial j_0}{\partial s_0}. \label{Lop}
\end{equation}

We can solve Equation \eqref{s1eq} with standard methods employed in stellar oscillation theory (or quantum mechanics; e.g., \citealt{hansen04}) and expand $\tilde{s}_1$ in terms of the eigenfunctions of $\mathcal{L}$, where the eigenfunctions $E_{\lambda}$ satisfy
\begin{equation}
\mathcal{L}\left[E_{\lambda}\right] = -\lambda^2E_{\lambda}. \label{eigens}
\end{equation}
The eigenvalues $\lambda^2$ are constrained by requiring that $E_{\lambda}$ satisfy the boundary conditions $E_{\lambda}(s_0 = 0) = 0$ (which it must by symmetry, i.e., the axis of the stream cannot be displaced for purely cylindrically symmetric perturbations) and $g_0E_{\lambda}(s_0 = 1) = 0$; the latter boundary condition enforces zero mass flux at the surface. The normalization of the eigenfunctions is arbitrary, and hence we can always let $E_{\lambda}(s_0 = 1) = 1$, meaning the system in general is over-constrained. The eigenfunctions satisfy all three boundary conditions, and since the operator given in Equation \eqref{Lop} can be put into Sturm-Liouville form and the eigenvalue equation is Hermitian, the eigenfunctions constitute a complete orthogonal basis (orthogonal with respect to the weight $s_0g_0$) that can be orthonormalized and the eigenvalues (strictly speaking $\lambda^2$) are purely real (see, e.g., \citealt{riley06}). A simple, brute-force method for computing the eigenvalues is to integrate Equation \eqref{eigens} from $s_0 = 0$ outward\footnote{Note that the series expansion of Equation \eqref{eigens} about the origin yields $\tilde{s}_1 \propto s_0$ to leading order in $s_0$, and the arbitrariness of the normalization implies that we can set $dE_{\lambda}/ds_0(s_0 = 0) = 1$ and renormalize the value at the surface to one after the eigenvalue is determined; in practice this is how we determine the eigenvalues and then normalize the eigenfunctions.} for an arbitrarily chosen value of $\lambda$. We then iterate on $\lambda$ until the boundary condition at the surface is satisfied to a high level of tolerance. Here we required the mass flux at the surface to be $< 10^{-10}$, and changing this criterion by an order of magnitude (in either direction) has no bearing on the eigenvalue to at least the fifth decimal place.  

Figure \ref{fig:eigenfuns} shows the first five eigenfunctions for a polytropic index of $\Gamma = 5/3$ (left) and $\Gamma = 4/3$ (right), and the eigenvalues are shown in the legend. The eigenfunctions exhibit the expected properties (see any book on quantum mechanics or the discussion of stellar pulsations in \citealt{hansen04}): there is a lowest-order mode that has no zero crossings (aside from the one at the origin), and each higher-order mode has one more zero crossing than the previous one. The lowest-order mode corresponds to purely outward or inward motion, and is the cylindrical analog of the ``breathing mode'' of (spherical) stellar oscillations. Higher-order modes, or ``overtones,'' have both inward and outward motion as a function of cylindrical radius at a given time, and are similar to p-modes in that most of the power is concentrated near the surface of the stream. 

Writing $\tilde{s}_1$ as a sum over the eigenfunctions
\begin{equation}
\tilde{s}_1 = \sum_{\lambda} c_{\lambda}(\sigma)E_{\lambda}(s_0),
\end{equation}
inserting this expansion into Equation \eqref{s1eq}, multiplying by $g_0 s_0$, integrating from $s_0 = 0$ to $s_0 = 1$ and using the orthogonality of the eigenfunctions, we find
\begin{equation}
c_{\lambda} = \frac{2F_{\lambda}}{2\sigma^2-\sigma+\mu^2\lambda^2}, \,\,\, F_{\lambda} = \frac{\int_0^{1}V_0E_{\lambda}g_0 s_0ds_0}{\int_0^{1}E_{\lambda}^2g_0s_0ds_0}. \label{clambda}
\end{equation}
The time-dependent solution is recovered by taking the inverse-Laplace transform, which from the residue theorem can be written as a sum over the poles of $c_{\lambda}$ in the complex plane; the result is
\begin{equation}
s_1(s_0,\tau) = \sum_{\lambda} E_{\lambda}(s_0)\frac{F_{\lambda}}{\sigma_{+}-\sigma_{-}}\left(e^{\sigma_{+}\tau}-e^{\sigma_{-}\tau}\right), \label{s1oft}
\end{equation}
where
\begin{equation}
\sigma_{\pm} = \frac{1}{4}\left(1\pm\sqrt{1-8\mu^2\lambda^2}\right). \label{sigmapm}
\end{equation}

To understand what this result implies about the stability of the stream, it is useful to first analyze the limiting case of $\mu = 0$: recall from Equation \eqref{mudef} that $\mu^2 \propto \rho_{\rm i}/\rho_{\bullet}$, so when $\mu = 0$, self-gravity and pressure are ignorable, and we expect the solution to be described by homologous expansion. With $\mu = 0$, $\sigma_{+} = 1/2$ and $\sigma_{-} = 0$, which shows that the stream width (from Equation \ref{spert}) satisfies
\begin{equation}
H(\tau) = H_{\rm i}e^{\tau/2}\left(1+C_1e^{\tau/2}\right),
\end{equation}
where $C_1$ is an arbitrary constant. This is just the exact solution to the equation of motion when only the tidal force acts to modify the stream width, as can be verified from Equation \eqref{smom1}, which is precisely what we expect. We therefore conclude that when the ratio of the stream density to the black hole density is small, the stream is unstable from the standpoint that it will transition to homologous expansion. 

As $\mu$ increases, $\sigma_{+}$ remains purely real and decreases in magnitude until $\mu$ reaches a critical value given by
\begin{equation}
\mu_{\rm cr}^2 = \frac{1}{8\lambda_0^2},
\end{equation}
where $\lambda_0 \simeq 2.669$ ($3.879$) is the smallest eigenvalue for $\Gamma = 5/3$ ($\Gamma = 4/3$), and thus $\mu_{\rm cr} \simeq 0.13$ ($\mu_{\rm cr} \simeq 0.091$). For this value of $\mu$ the $\lambda_0$ term in the sum in Equation \eqref{s1oft} is a repeated root at $\sigma = 1/4$ (it was assumed in deriving this equation that $\sigma_{+} \neq \sigma_{-}$), and the application of the residue theorem (or taking the limit that $\sigma_{-} \rightarrow \sigma_{+}$ in Equation \ref{s1oft}) for the repeated root shows that the $\lambda_0$ solution is $\propto \tau e^{\tau/4}$; recalling that $\tau \propto \ln R_{\rm c}$, we see that for $\mu = \mu_{\rm cr}$ the width of the stream expands as
\begin{equation}
H(t) \propto \tau e^{\tau/4} \propto t^{1/6}\ln t,
\end{equation}
where we used the fact that $R_{\rm c} \propto t^{2/3}$. From Equation \eqref{mudef}, this value of $\mu$ corresponds to a ratio of stream to black hole density of
\begin{equation}
\frac{\rho_{\rm cr}}{\rho_{\bullet}}(\Gamma = 5/3) \simeq 0.102, \quad \frac{\rho_{\rm cr}}{\rho_{\bullet}}\left(\Gamma = 4/3\right) \simeq 0.148.
\end{equation}
Here we used the value of $\alpha_{\rm i}$ appropriate to each polytropic index, as given by Equation \eqref{Kstar}. For this critical value of $\mu$, the stream is therefore unstable and grows as $\propto t^{1/6}$ with an added, logarithmic boost to the growth rate.

For $\mu > \mu_{\rm cr}$, $\sigma_{+}$ and $\sigma_{-}$ are complex with a real part of $1/4$ (see Equation \ref{sigmapm}), and the temporal evolution of each term in the series expansion is, from Equation \eqref{s1oft},
\begin{equation}
e^{\tau/4}\sin\left(\frac{\tau}{4}\sqrt{8\mu^2\lambda^2-1}\right) \propto t^{1/6}\sin\left(\frac{\ln t}{6}\sqrt{8\mu^2\lambda^2-1} \right). \label{aux1}
\end{equation}
In the last expression time is measured relative to the dynamical time at the scale distance $R_{\rm i}$, which should be comparable to the dynamical time at the tidal radius and is, by construction, also equal to the dynamical time of the star. Because the oscillatory nature of the solutions proceeds logarithmically with time, it follows that the oscillation period is {exponentially long}. 

We interpret $\rho_{\rm cr}$ as the minimum density necessary for the steam to retain approximate hydrostatic balance, while those with $\rho_{\rm i}/\rho_{\bullet} < \rho_{\rm cr}$ expand $\sim$ ballistically in the tidal field of the black hole. In support of this interpretation, note from Equation \eqref{aux1} that in the limit of $\mu \gg 1$ the oscillation frequencies scale as $\sim \lambda \mu$ and the dominant behavior of the stream is to oscillate on the dynamical timescale of the stream. The growth rate of the amplitude of the oscillations is very small relative to the dynamical time of the stream in this limit, because the growth rate depends only on $\tau$ while the oscillation frequency is increased by $\mu$. On the other hand, since the lowest-order mode is monotonically increasing with cylindrical radius and approximately homologous (see Figure \ref{fig:eigenfuns}), if $\sigma$ is purely real the stream will simply continue to expand forever until $\sim$ homologous expansion (with $H \propto e^{\tau}$) is reached; this is the dominant solution to the purely dynamical equation -- Equation \eqref{smom1} without the pressure or self-gravity terms.

We note that Equation \eqref{sigmapm} is similar to what was proposed by \citet{coughlin16} (see their Section 6.3) to characterize the oscillations of the fluid. In particular, by reasoning analogously to galaxy formation in the early (expanding) Universe, they argued that the modes would be power-laws in time, and that the growth timescale would be proportional to $\sqrt{\rho_{\rm i}/\rho_{\bullet}}$. We see that the first of these predictions is correct (recall that $e^{\sigma\tau} \sim R_{\rm c}^{\sigma} \sim t^{2\sigma/3}$), while the second actually only applies to the oscillation frequency of the stream when $\mu \gg 1$. On the other hand, when $\mu$ is sufficiently small the frequency no longer scales in this way. The growth rate of the overstability is also $\propto t^{1/6}$, and is independent of the mass ratio $\rho_{\rm i}/\rho_{\bullet}$. 

In the next section we consider a specific example of how this instability (and overstability) operates in TDEs.

\subsection{An Example}
As an example, both numerical simulations \citep{guillochon13, mainetti17, miles20} and analytical analyses (\citealt{coughlin22a} and the discussion in Section \ref{sec:basic}) have found that the tidal disruption radius of a $5/3$, polytropic star occurs at $R_{\rm t} \simeq R_{\star}\left(M/M_{\star}\right)^{1/3} = R_{\rm i}$, and hence $\rho_{\bullet} \simeq \rho_{\star}$, where $\rho_{\star}$ is the average stellar density. A $5/3$ polytrope also has $\rho_{\rm i}/\rho_{\star} \simeq 6$, and hence (from Equation \ref{mudef})
\begin{equation}
\mu = \sqrt{\frac{3\rho_{\rm i}}{\rho_{\star}}\alpha_{5/3}} \simeq 1.01.
\end{equation}
This is a factor of $\sim 10$ greater than the critical $\mu$ that delimits pure instability from overstability, and hence the stream radius as a function of time that includes perturbations from the lowest-order mode -- which are most likely to be largest in terms of the initial perturbations present on the stream, e.g., the ensuing example in which we consider a homologous initial velocity perturbation -- scales as
\begin{equation}
H(\tau) = H_{\rm i}e^{\tau/2}\left\{1+\delta H \left(\frac{R_{\rm c}}{R_{\rm t}}\right)^{1/4}\sin\left[1.9\ln\left(\frac{R_{\rm c}}{R_{\rm t}}\right)\right]\right\}. \label{Hoftau}
\end{equation}
This expression comes from combining Equation \eqref{spert} with the solution for the perturbation to the lowest-order mode (Equation \ref{aux1}), noting that the surface coincides with $s_0 = 1$, and $\delta H$ is the magnitude of initial perturbation to the stream that arises from the fundamental mode. The numerical factor of 1.9 is equal to $1/6\times\sqrt{8\mu^2\lambda_0^2-1}$ with $\mu = 1.01$ and $\lambda_0 = 2.669$. We see that the oscillations occur on exponentially long timescales, such that the $n$th oscillation occurs when 
\begin{equation}
\frac{R_{\rm c}}{R_{\rm t}} = e^{2\pi n/1.9},
\end{equation} 
or using the expression for the marginally bound Keplerian orbit as a function of time, on timescales
\begin{equation}
T_{\rm n} = \frac{R_{\rm t}^{3/2}}{\sqrt{2GM_{\bullet}}}e^{3\pi n/1.9} \simeq t_{\star}e^{5n}, \label{Tn}
\end{equation}
where $t_{\star} = R_{\star}^{3/2}/\sqrt{2GM_{\star}}$ is roughly the dynamical time of the star. For a solar-like polytrope, the first complete oscillation ($n = 1$) occurs after $\sim 1.9$ days, the second oscillation after $\sim 287$ days, and the third on a timescale of $\sim 117$ years.

As a second example, the zero-age main sequence Sun is well-approximated by the Eddington standard model, and simulations and analytical estimates have found that this type of star is tidally destroyed when the pericenter distance of the center of mass comes within $R_{\rm p} \simeq R_{\rm t}/1.8$. For a $4/3$ polytrope, $\rho_{\rm i}/\rho_{\star} \simeq 54$, and hence we have
\begin{equation}
\mu = \sqrt{\frac{3\rho_{\rm i}}{\rho_{\star}\left(1.8\right)^3}\alpha_{4/3}} \simeq 0.740,
\end{equation}
which is a factor of $\sim 5$ times the critical density for overstability. Following the same steps as we did for the $5/3$ polytrope, the oscillation periods for this type of star are
\begin{equation}
T_{\rm n} \simeq t_{\star}e^{4.7 n},
\end{equation}
and thus the first oscillation occurs on $\sim 1.4$ days, the second on $\sim 152$ days, and the third on $\sim 45$ years. 

Tidally disrupted debris streams that satisfy $\rho_{\rm i} > \rho_{\rm cr}$ are therefore appropriate for most stars (both radiative and convective) that are destroyed by SMBHs, and are therefore quasi-stable from the standpoint that self-gravity is able to confine the stream, but the time-dependence of the background (expanding) gas results in the system overshooting its equilibrium and oscillating with a growing amplitude. Therefore, it seems likely that once the amplitude of the perturbation grows to become of the order unity, then the next oscillation that results in a relative maximum of the perturbation will cause the stream to ``bounce'' out of equilibrium and approach homologous expansion (see Appendix \ref{sec:homologous} and Figure \ref{fig:deltaH}). Because the amplitude grows as a power-law in time and with a small power-law index, the time taken for this condition to be reached can be extremely long. Setting $\delta H\left(R_{\rm c}/R_{\rm t}\right)^{1/4} = 1$ in Equation \eqref{Hoftau}, the position of the marginally bound radius at which the magnitude of the perturbation is comparable to 1 and hence at which this is expected to occur, which we define as $R_{\rm c, b}$, and the time at which this should occur, which we denote $t_{\rm b}$, are
\begin{equation}
R_{\rm c, b} = R_{\rm t}\left(\frac{H_0}{\delta H}\right)^{4} \quad \Rightarrow \quad t_{\rm b} \simeq t_{\star}\left(\frac{H_0}{\delta H}\right)^{6},
\end{equation}
where $H_0$ is the unperturbed stream width. 

The timescale taken for the stream to bounce out of equilibrium clearly depends sensitively on the magnitude of $\delta H$. One mechanism that is responsible for inducing perturbations on the stream is the in-plane pancake discussed in \citet{coughlin16b}. In particular, the orbital motion of the fluid as the stellar center of mass passes through pericenter results in the dynamical focusing of the stream within the orbital plane of the star and a convergence of the fluid. If the convergence of the fluid is approximately homologous in terms of the velocity, such that the initial (dimensionless, i.e., in units of $H_{\rm i}/t_{\star}$) velocity of the fluid elements is $v_0 = V_{\rm i}s_0$ with $V_{\rm i}$ a constant, then the coefficients of the eigenmode expansion, $F_{\lambda}$, can be computed from Equation \eqref{clambda} with this specific velocity profile and inserted into Equation \eqref{s1oft} to determine the time-dependent solution.

\begin{figure}
   \centering
   \includegraphics[width=0.47\textwidth]{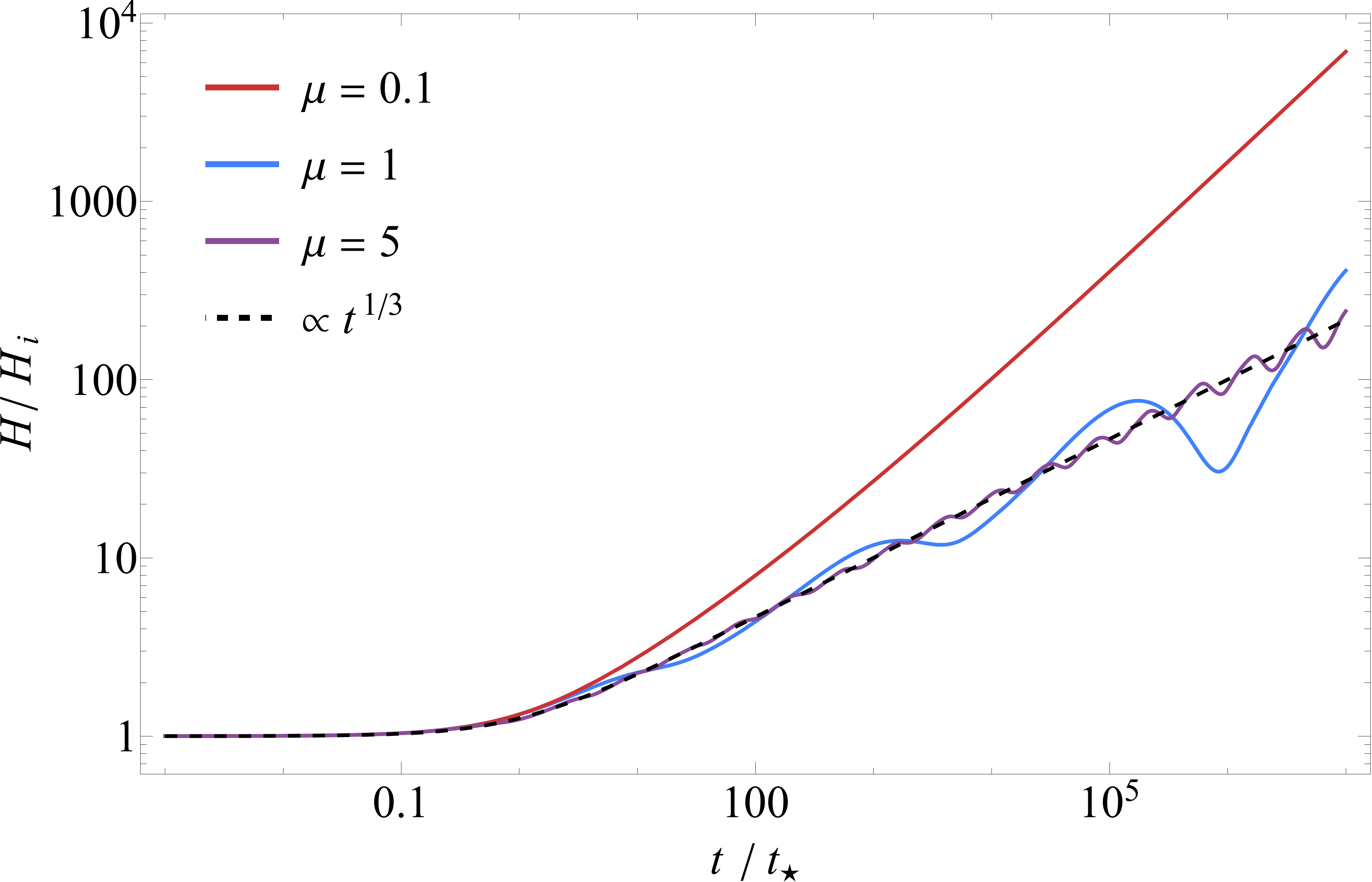} 
    \caption{The evolution of the cylindrical radius of the stream, $H/H_{\rm i}$ with $H_{\rm i}$ the radius at $t = 0$, as a function of time in units of the dynamical time of the star; here an initial, homologous velocity profile with a magnitude of $V_{\rm i} = 0.1$ provides the initial perturbation. The value of $\mu$ is shown in the legend; $\mu = 0.1$ is unstable, and $\mu = 1$ and 5 are both overstable, with an oscillation frequency that scales approximately with $\mu$. The black, dashed line shows the unperturbed solution for reference.}
   \label{fig:H_of_t}
\end{figure}

\begin{figure*}
   \centering
   \includegraphics[width=0.47\textwidth]{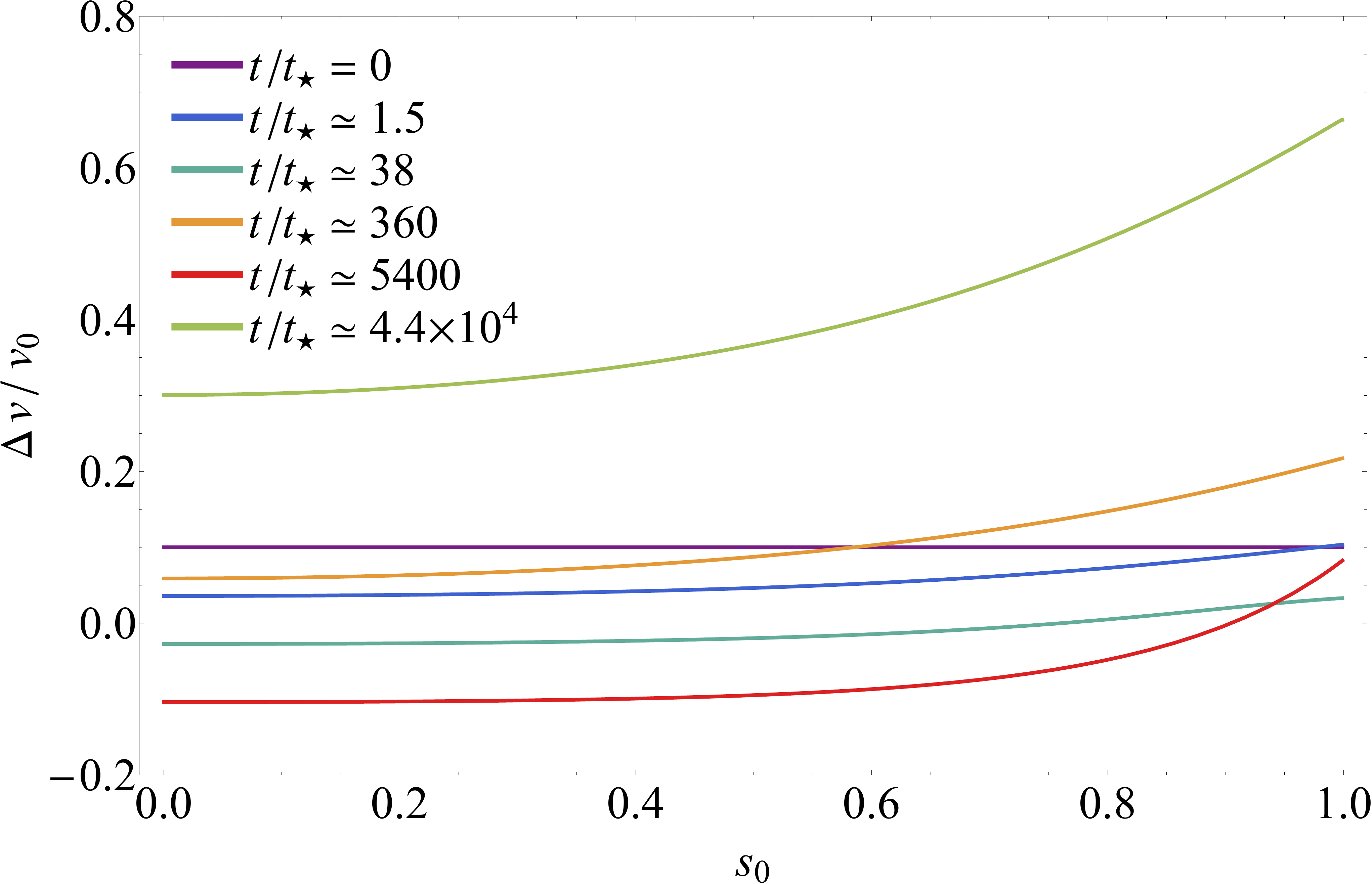} 
    \includegraphics[width=0.47\textwidth]{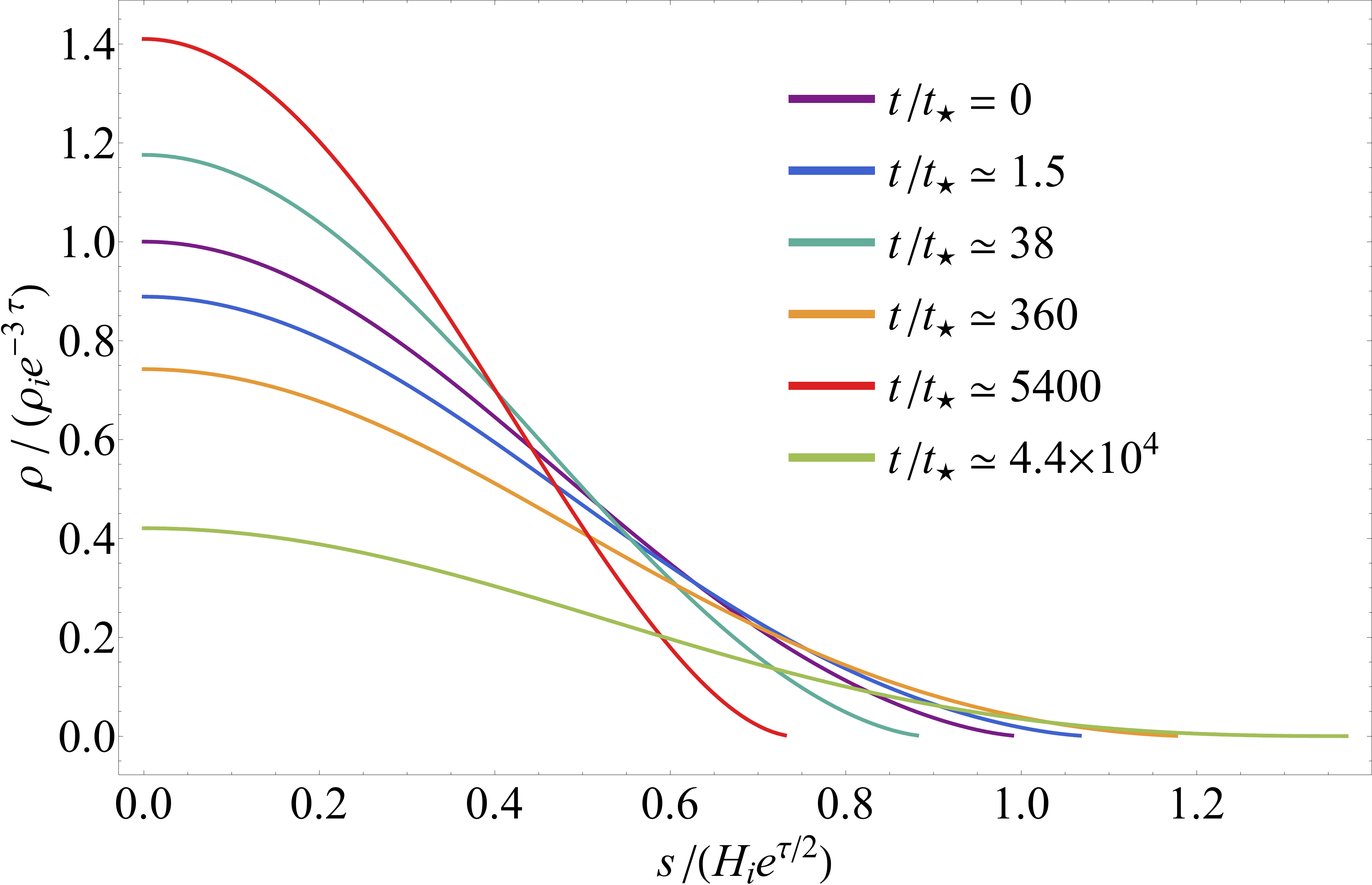} 
   \caption{Left: The ratio of the velocity difference (including the perturbations) to the unperturbed velocity within the stream as a function of the initial Lagrangian radius, $s_0$. The perturbation in this case is a homologous initial velocity with magnitude $V_0 = 0.1$ (see Equation \ref{s1pert}) and $\mu = 1$. The legend gives the time (in units of $T_{\star} = R_{\star}^{3/2}/\sqrt{2GM_{\star}}$) at which the velocity profile is measured. Note that since the unperturbed velocity is $\propto s_0$, the solution at $t = 0$ is just a horizontal line at the magnitude of the perturbation, which in this case is 0.1. Right: The density of the stream normalized by $\rho_{\rm i}e^{-3\tau}$, which is the background, overall temporal scaling of the unperturbed solution, as a function of Eulerian cylindrical radius $s$ normalized by the scaling of the background solution $H_{\rm i}e^{\tau/2}$. }
   \label{fig:Deltav}
\end{figure*}

Figure \ref{fig:H_of_t} shows the evolution of the stream width for the $\mu$ shown in the legend when $V_{\rm i} = 0.1$ (the dashed line gives the temporal scaling of the unperturbed solution). The gas is polytropic ($K_0 \equiv 1$) in this case, and we used the first 10 eigenmodes. When $\mu = 0.1$, the solution is just below the critical value necessary to be overstable, and the stream radius monotonically diverges from the background solution. When $\mu = 1$, which is representative of the value likely to be realized in most tidal disruption events, the stream radius oscillates increasingly violently and over/undershoots the equilibrium by a factor of the order unity by $\sim 10^{5}$ dynamical times. For $\mu = 5$, the solution oscillates many times (as all of the frequencies are increased by a factor of $\sim 5$ relative to $\mu = 1$) and still represents a relatively small perturbation by $10^6$ dynamical times.

The left panel of Figure \ref{fig:Deltav} shows the difference between the perturbed velocity (where the velocity is $v = \partial s/\partial t$) and the unperturbed velocity, normalized by the unperturbed velocity, as a function of the initial Lagrangian position and for the times shown in the legend. The right panel of this figure shows the density, normalized by the temporal scaling of the background solution, as a function of the current Lagrangian position (i.e., this is the Eulerian density profile) relative to the background scaling of the stream width. Here we set $\mu = 1$ and $V_{\rm i} = 0.1$. Since the initial perturbation is a homologous velocity profile and the unperturbed velocity is also homologous, at $t = 0$ the solution for the velocity is a constant and equal to $V_{\rm i}$ (the fact that the solution recovers this is actually a check on the accuracy of our eigenmode decomposition). As time advances, the velocity and the density oscillate about their initial values and increasingly violently, and by $\sim 4.4\times 10^{4}$ dynamical times, the relative difference in the velocity approaches $\sim 1$, signaling the breakdown of the perturbation method.

\begin{figure} 
   \centering
   \includegraphics[width=0.475\textwidth]{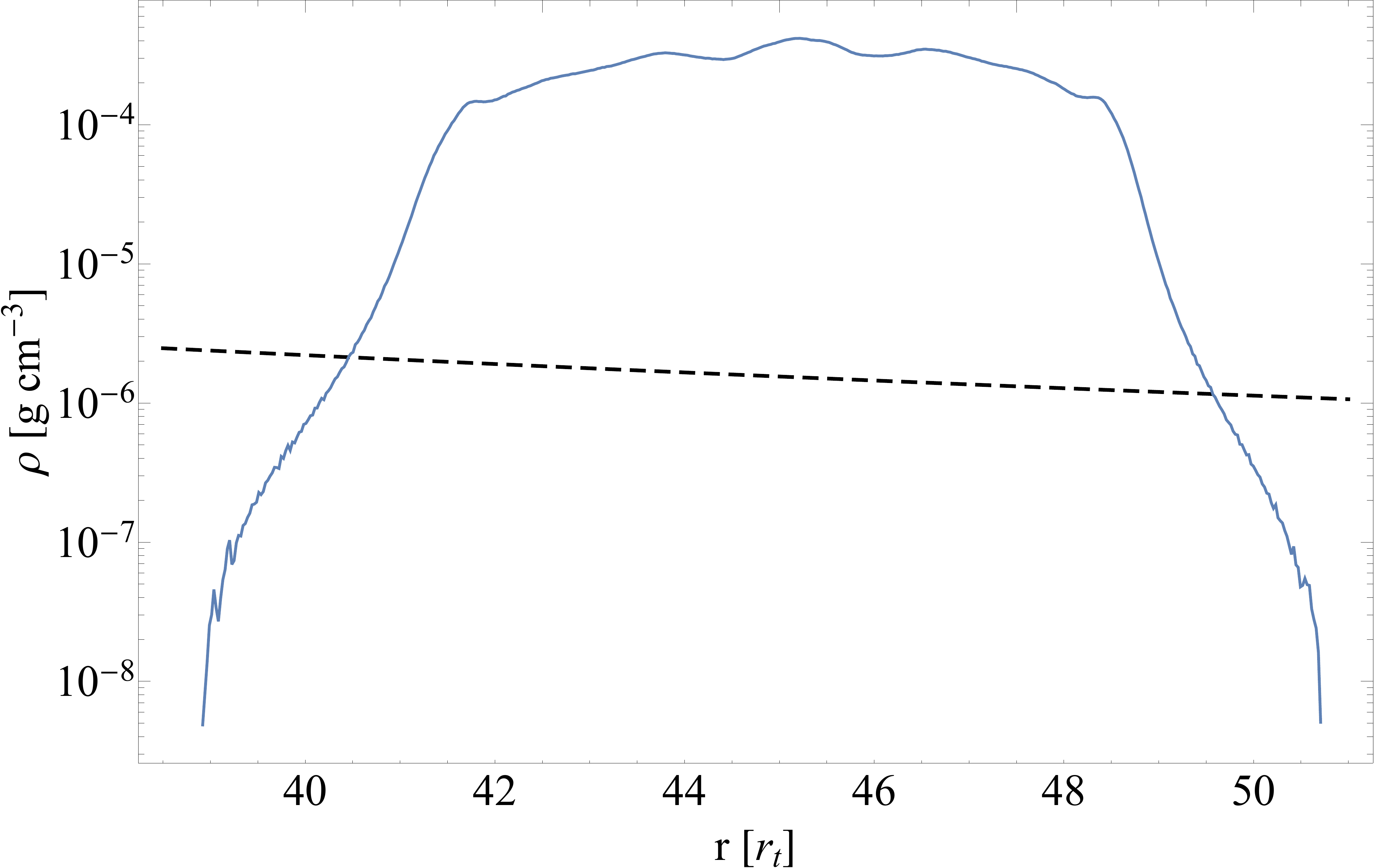} 
   \caption{The density of the debris stream produced from the tidal disruption of a solar-like, 5/3 polytrope by a $10^6M_{\odot}$ SMBH, as a function of distance from the SMBH in units of $r_{\rm t} = 100R_{\odot}$. The black, dashed line shows the critical stream density, below which we do not expect the stream to be self-gravitating. The density is averaged over the small solid angle subtended by the stream (see Figure \ref{fig:debris_stream}). }
   \label{fig:rho_of_r}
\end{figure}

\begin{figure} 
   \centering
   \includegraphics[width=0.4\textwidth]{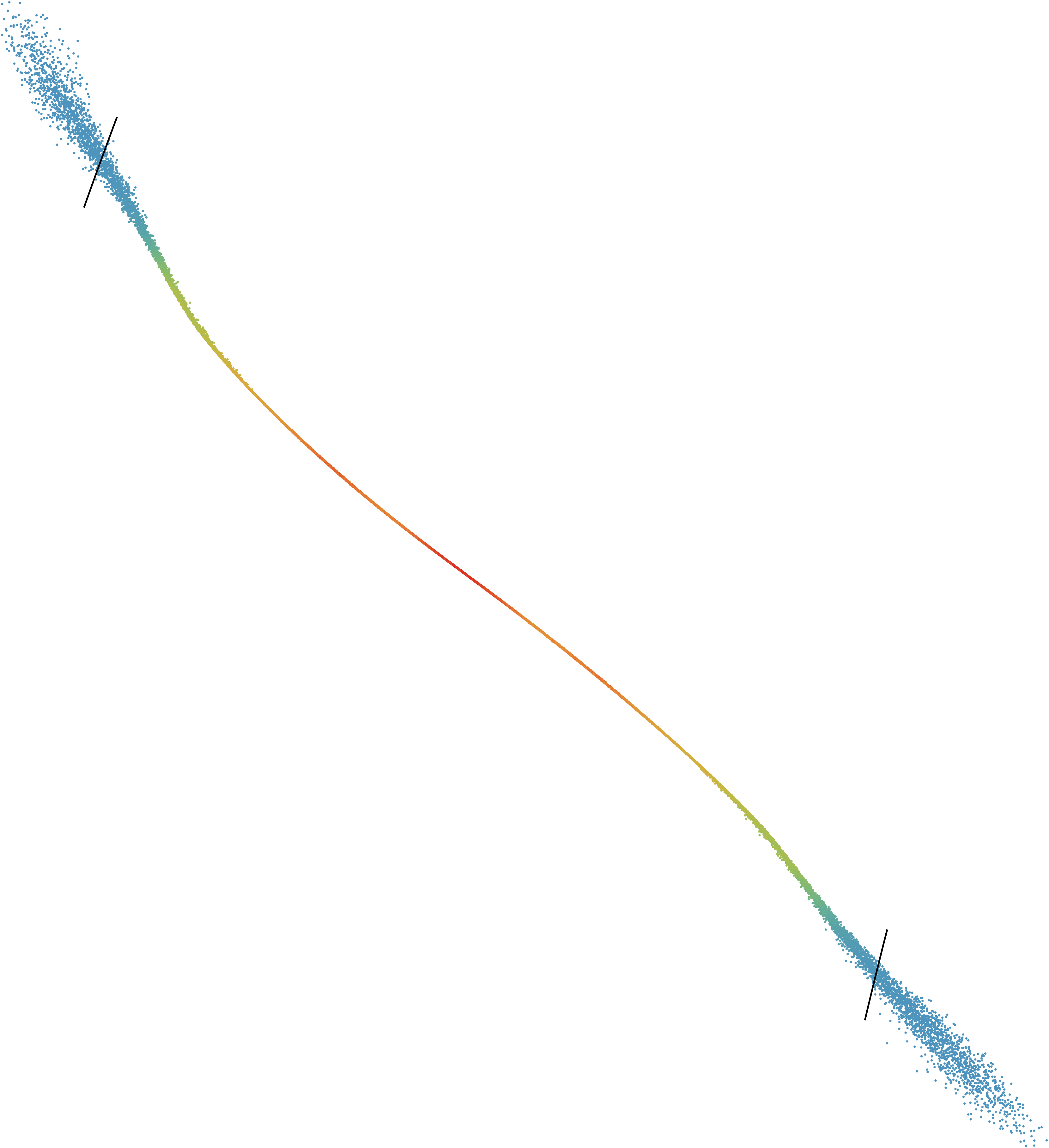} 
   \caption{A subset of the particles within the debris stream at the same time as Figure \ref{fig:rho_of_r} from the same hydrodynamical simulation, where the colors scale with the log of the density (red is densest, blue is least dense). Particles to the left of the top-left, black line or to the right of the bottom-right, black line fall below the critical density to be self-gravitating. These radii coincide closely with where the debris stream ``fans out'' and is geometrically noticeably thicker. The SMBH is to the right in this figure.}
   \label{fig:debris_stream}
\end{figure}

In a tidally disrupted debris stream from a TDE, there is a gradient in the density along the axis that results from the fact that the core of the star had the highest density. Therefore, we would expect the location along the stream where the density satisfies $\rho/\rho_{\bullet} \simeq 0.1$ to coincide with where self-gravity no longer confines the material. Figure \ref{fig:rho_of_r} shows the density along the debris stream produced from the ``canonical TDE'' -- a solar-like, 5/3 polytrope destroyed by a $10^6M_{\odot}$ SMBH with a pericenter distance of $r_{\rm t} = 100R_{\odot}$ -- simulated with the SPH code {\sc phantom} \citep{price18}. We used $\sim 10^6$ particles and the equation of state is adiabatic with $\gamma = 5/3$, and the reader is referred to \citet{coughlin15, price18} for additional details of the setup and the self-gravity solver. The black, dashed curve shows $\rho = 0.1 \rho_{\bullet}$, and thus the regions of the stream that are below this density should not be self-gravitating. Figure \ref{fig:debris_stream} shows a subset of the SPH particles, where the colors scale with the base-10 logarithm of the density (red is highest density, blue is lowest density). The black lines are where the density falls below the critical density to be self-gravitating, and we see that this location coincides closely with where the stream ``fans out,'' and becomes noticeably thicker relative to the geometric center (the marginally bound radius in this figure occurs at $\sim 45.2 r_{\rm t}$). The critical density thus characterizes the location along the stream where the stream goes from narrow and gravitationally confined to wider and non-self-gravitating.

\begin{figure*}
   \centering
   \includegraphics[width=0.495\textwidth]{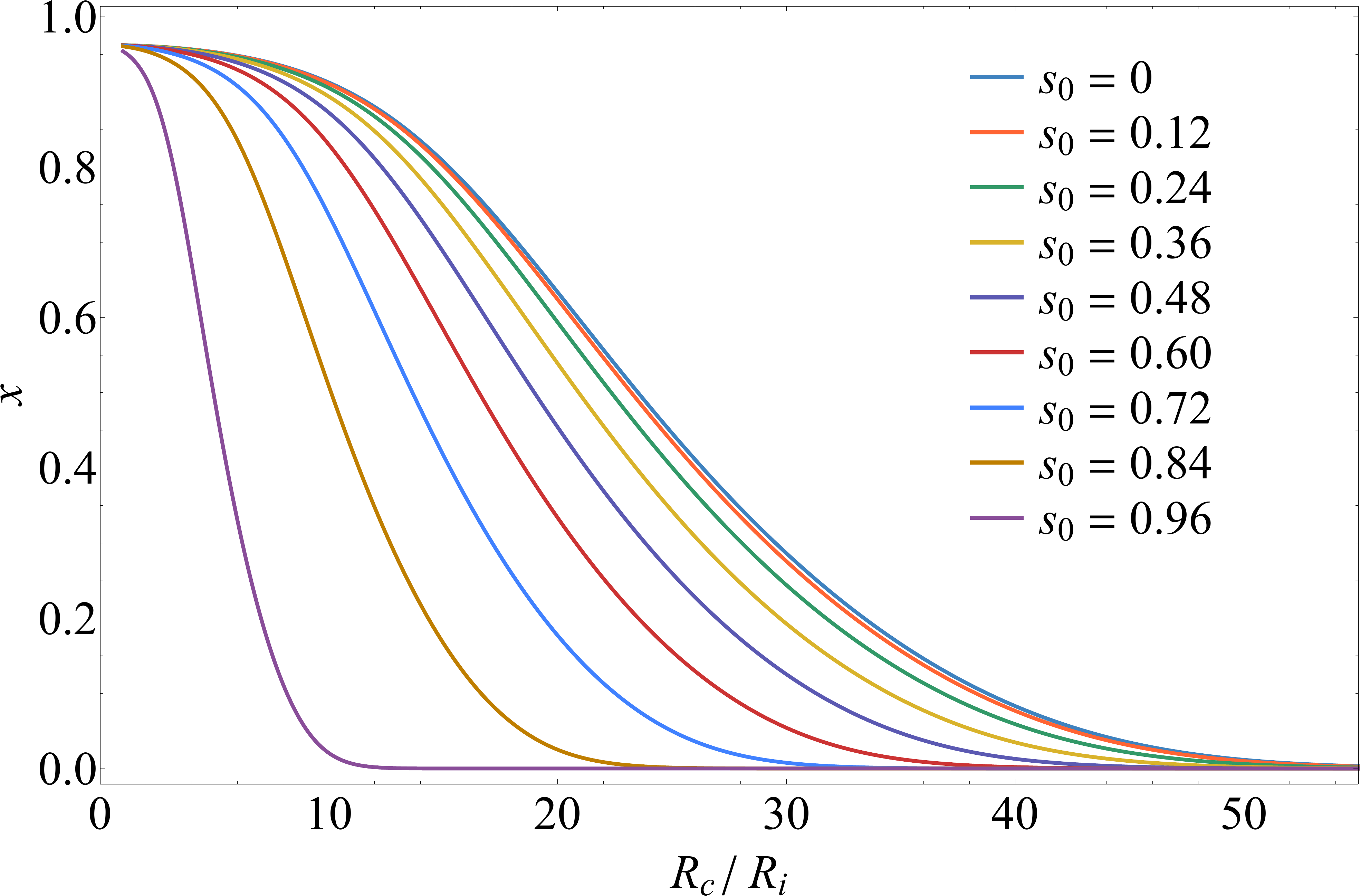} 
 \includegraphics[width=0.495\textwidth]{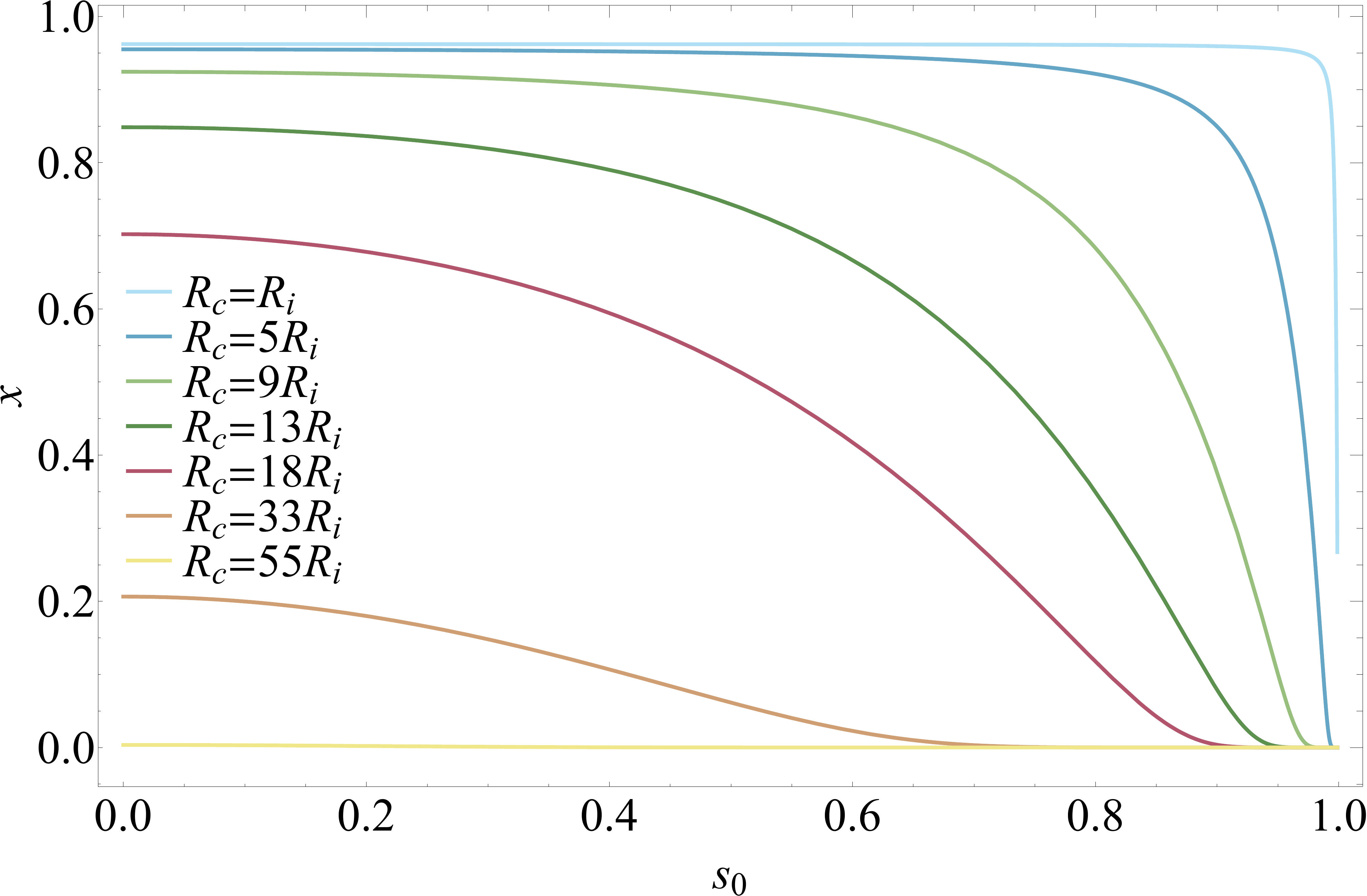} 
   \caption{Left: The hydrogen ionization fraction (i.e., the ratio of the number of free electrons to the number of ionized and neutral hydrogen atoms) as a function of the position of the Keplerian zero-energy orbit. The different curves are for the different initial cylindrical radii within the stream; because the initial temperature is lower near the surface of the stream, these regions recombine earlier. Right: The hydrogen ionization fraction as a function of the initial cylindrical radius within the stream for the times positions of the Keplerian zero-energy orbit within the legend (for reference, $R_{\rm c}/R_{\rm i} \simeq 1.5$ days after disruption for a sun-like star). This figure demonstrates, analogously to the left panel, that recombination happens predominantly from the outside-in, and by $R_{\rm c}/R_{\rm i} \simeq 50$, the entire stream has recombined (consistent with the left panel).}
   \label{fig:xofRc}
\end{figure*}

In the next section we investigate an additional perturbation that modifies the stream, which is the recombination of hydrogen. For concreteness we assume that the background state is polytropic for the remainder of the paper.

\section{hydrogen recombination}
\label{sec:recombination}
A thermodynamic effect that has relevance for the evolution of the debris stream is when the gas cools to the point that it starts to recombine, and we can use the analytic solutions so-far obtained to understand the impact of recombination on the stream. From the exact solution above the temperature\footnote{Radiation pressure is ignorable because the ratio $T^{3}/\rho$ declines with time, and for all stars that are of relatively low-mass, radiation pressure is insignificant in the stellar interior.} of the gas within the expanding debris stream is
\begin{equation}
T \propto \frac{p}{\rho} = T_{\rm i}e^{-2\tau} = T_{\rm i}\left(\frac{R_{\rm c}}{R_{\rm i}}\right)^{-2} \propto t^{-4/3}, \label{Tgas}
\end{equation}
If we focus on the material near the Keplerian marginally bound radius and that contains most of the mass, then the initial temperature (for a Sun-like star) is $T_{\rm i} \simeq 10^{7}$ K, and the temperature of the gas will fall to $\sim 10^{4}$ K when the center of mass reaches
\begin{equation}
R_{\rm rec} \simeq 30 R_{\rm i} \quad \Rightarrow \quad t_{\rm rec} \simeq 84 t_{\star},
\end{equation}
where $t_{\star} = R_{\star}^{3/2}/\sqrt{GM_{\star}}$ is the dynamical time of the star (as also introduced in the previous section) and we assumed the initial position of the center of mass was equal to the tidal radius. For a sun-like star with $R_{\star} = 1R_{\odot}$ and $M_{\star} = 1M_{\odot}$, the recombination timescale is $t_{\rm rec} \sim 1.5$ days. 

Once the gas falls below $\sim 10^{4}$ K we expect hydrogen recombination to occur and modify the thermodynamics of the stream. We can assess recombination quantitatively by noting that, assuming thermodynamic equilibrium and (hence) that the recombination timescale is short enough that dynamical expansion on that timescale is ignorable, the Saha equation,
\begin{equation}
\frac{x^2}{1-x} = \frac{m_{\rm H}}{\rho}\left(\frac{2\pi m_{\rm e}kT}{h^2}\right)^{3/2}e^{-\epsilon_{\rm H}/kT} \label{saha},
\end{equation}
can be used to determine the hydrogen ionization fraction. Here $\epsilon_{\rm H} = 13.6$ eV is the ionization energy of hydrogen, $m_{\rm e}$ and $m_{\rm H}$ are the electron and hydrogen mass, $\rho = m_{\rm H}\left(n_{\rm p}+n_{0}\right) \equiv m_{\rm H}n$ with $n_{\rm p}$ and $n_{0}$ the number density of ionized and neutral hydrogen, respectively, and $x = n_{\rm p}/n$ is the hydrogen ionization fraction. In this equation, we can use the ideal gas law,
\begin{equation}
p = \left(1+x\right)\frac{\rho k T}{m_{\rm H}},
\end{equation}
to relate the temperature to the pressure and density, both of which can be approximated from the exact solution in the adiabatic limit (while this is only an approximation, it gives a useful estimate for understanding how recombination proceeds in the stream given this background state), and the ionization fraction $x$, and inserting the result into Equation \eqref{saha} allows us to solve (numerically) for the ionization fraction as both a function of time and initial cylindrical radius within the stream (note that the temperature $T$ is also solved for alongside the ionization fraction). 

The left panel of Figure \ref{fig:xofRc} shows the hydrogen ionization fraction as a function of $R_{\rm c}/R_{\rm i}$ that results from solving the Saha equation for the cylindrical radii shown in the legend. Here we chose an initial gas temperature of $T_{\rm i} \simeq 10^{7}$ K and an initial density of $\rho_{\rm i} = 10$ g cm$^{-3}$. This figure demonstrates that, consistent with the estimate above, hydrogen recombination starts to occur vigorously along the axis of the stream after the zero-energy Keplerian orbit expands to $\sim 20-30$ times its initial position. However, this figure also shows that hydrogen recombination occurs from the outside in -- fluid elements at larger initial radii (with lower initial temperatures) fall below $\sim 10^{4}$ K sooner, and thus recombine sooner as well. This figure also shows that, while the majority of the stream has recombined by $R_{\rm c}/R_{\rm i} \simeq 30$, in agreement with the estimate above, the process starts somewhat sooner and finishes completely by $R_{\rm c}/R_{\rm i} \simeq 50$. The right panel of this figure shows the ionization fraction as a function of cylindrical radius for the $R_{\rm c}/R_{\rm i}$ in the legend. Again, this demonstrates that recombination occurs from the outside-in, and is effectively complete by a time of $R_{\rm c}/R_{\rm i} \simeq 50$, which corresponds to $\sim 2.5$ days for the disruption of a solar-like star.

\subsection{Recombination transient}
\label{sec:transient}
As the stream recombines, a fraction of the energy may be radiated from the system and reach the observer, resulting in a ``recombination transient'' that has been analyzed by \citet{kasen10}. The maximum possible energy able to be radiated per unit time is equal to the ionization potential multiplied by the total number of recombinations that occur during that time, i.e., if none of the recombination energy is trapped within the flow, which is given by
\begin{equation}
L_{\rm rec} = \frac{d}{dt}\left[\epsilon_{\rm H}\int\left(1-x\right)ndV\right] = -\epsilon_{\rm H}\frac{d}{dt}\int nxs\,ds\,dz\,d\phi.
\end{equation}
The number density is given by $\rho/m_{\rm H}$, from mass conservation $\rho s ds dz d\phi = \rho_0(s_0) s_0 ds_0 dz_0 d\phi_0$, and changing variables from $t$ to $\tau$ yields
\begin{equation}
\begin{split}
L_{\rm rec} &= -\frac{\epsilon_{\rm H}}{m_{\rm H}}\frac{V}{R}\frac{d}{d\tau}\int x\rho_0(s_0)s_0\,ds_0\,dz_0\,d\phi_0 \\
&= -\frac{\epsilon_{\rm H}}{m_{\rm h}}\frac{M_{\star}\sqrt{2GM_{\bullet}}}{R_{\rm c}^{3/2}}\frac{d}{d\tau}\int_0^{1}xg_0(s_0)s_0\,ds_0 \\
&= -\frac{\epsilon_{\rm H}M_{\star}\sqrt{2GM_{\bullet}}}{m_{\rm H}R_{\rm i}^{3/2}}\left(1+\frac{3}{2}\frac{\sqrt{2GM_{\bullet}}}{R_{\rm i}^{3/2}}t\right)^{-1} \int_0^{1} g_0(s_0)s_0\frac{dx}{d\tau}ds_0. \label{Lrec}
\end{split}
\end{equation}
In the last equality we used Equation \eqref{Rcoft} to write $R$ as a function of time, and the factor of $M_{\star}$ results from requiring that the total mass be equal to the mass of the star (i.e., if $x = 1$, then the integral over the ionization fraction must equal the mass of the star).

\begin{figure} 
   \centering
   \includegraphics[width=0.475\textwidth]{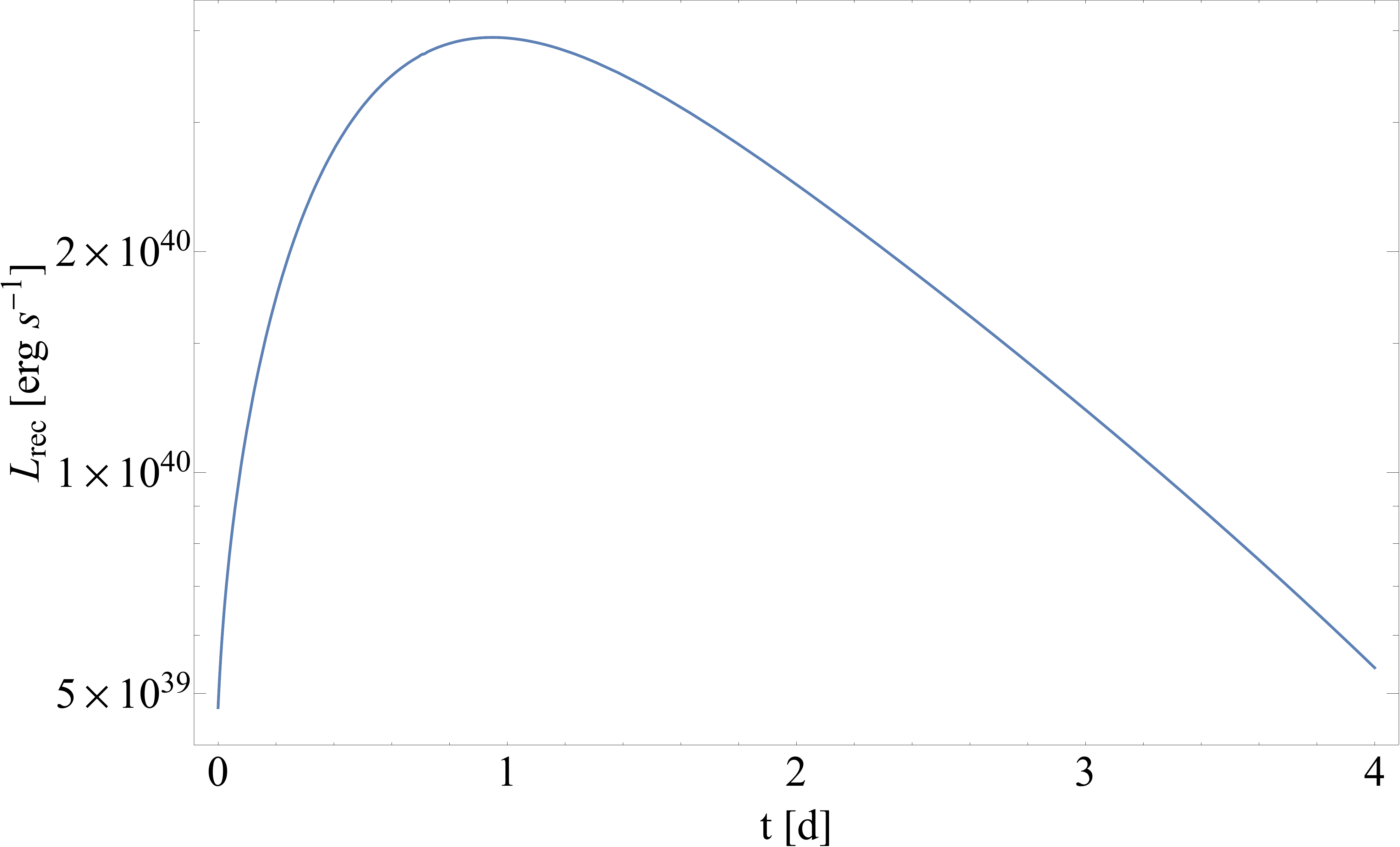} 
   \caption{The luminosity of the recombination transient as a function of time in days, calculated under the very optimistic assumption that the energy from recombination could be radiated promptly. The disrupted star in this case was assumed to have a solar mass and radius.}
   \label{fig:Lrec}
\end{figure}

Figure \ref{fig:Lrec} shows the recombination luminosity as a function of time in days that results from Equation \eqref{Lrec}. Here we assumed that $R_{\rm i} = R_{\star}\left(M_{\bullet}/M_{\star}\right)^{1/3}$ and that the star had a solar mass and radius. In this extremely optimistic case, the recombination transient reaches a peak magnitude of $\sim 4\times 10^{40}$ erg s$^{-1}$. This value for the peak luminosity is, if the disrupting SMBH has a mass of $10^{6}M_{\odot}$, approximately 5 orders of magnitude below the accretion luminosity from the fallback of the debris, being $L_{\rm fb} \simeq 5\times 10^{45}$ erg s$^{-1}$ if the radiative efficiency associated with accretion is 0.1 (see, e.g., Figure 3 of \citealt{coughlin15}).

The peak magnitude obtained in Figure \ref{fig:Lrec} is comparable to or somewhat larger than the peak luminosity deduced by \citet{kasen10} (see specifically their Figure 4). The fact that our value is somewhat larger is almost certainly related to the efficiency with which we assumed the recombination energy could be lost from the system -- we assumed that the energy could be radiated immediately, whereas \citet{kasen10} argued that the debris would have to cool to the point that the opacity was low enough that the material would be optically thin. This condition, they argued, would be when the gas fell to a temperature of $\sim 5,000$ K, at which the opacity of the gas reached a relative minimum. \citet{kasen10} also included more realistic radiative transfer calculations. Additionally accounting for the fact that the temperature will decline much more slowly during the recombination phase (see the next subsection), this will significantly lengthen the amount of time over which the energy is radiated and correspondingly reduce the luminosity in Figure \ref{fig:Lrec}. 

On the other hand, \citet{kasen10} made a number of assumptions about the nature of the debris stream to calculate their lightcurves that are likely not realistic for the vast majority of TDEs (or even any). For one, they assumed that the debris stream properties were largely homogenized due to the passage of a strong shockwave through the gas near the pericenter of the stellar orbit, which was motivated by the work of \citet{carter83} for deep TDEs (those in which the center of mass of the star reaches $\beta \gtrsim 3$, where $\beta = r_{\rm t}/r_{\rm p}$ and $r_{\rm p}$ is the pericenter distance of the star). It has since been demonstrated that such strong shocks do not exist, even for extremely deep (and rare) encounters with $\beta \gtrsim 10$ \citep{norman21, coughlin22b}. They also assumed that the debris expands homologously, again motivated by the passage of a strong shock that would eject the material, while our work here (and that of \citealt{kochanek94}) demonstrates that this is not the case -- the stream can maintain rough hydrostatic balance in its transverse directions. For this reason, the density and the optical depth remain much higher than what was predicted by \citet{kasen10}; specifically, we find that at $R_{\rm c}/R_{\rm i} = 30$ (where the temperature drops to $10^{4}$ K), the density has fallen by a factor of $\sim 30^{-3} \simeq 3.7\times 10^{-5}$, while the stream radius has expanded by a factor of $\sim 30^{1/3} \simeq 5.5$, and hence the optical depth across the stream is approximately
\begin{equation}
\begin{split}
\tau &\simeq \kappa \rho H = \kappa\rho_{\rm i}H_{\rm i}\left(\frac{R_{\rm c}}{R_{\rm i}}\right)^{-5/2} \\
&\simeq 5\times 10^{7} \left(\frac{\kappa}{0.34\textrm{ cm}^2\textrm{ g}^{-1}}\right)\left(\frac{\rho_{\rm i}}{10 \textrm{ g cm}^{-3}}\right)\left(\frac{H_{\rm i}}{R_{\odot}}\right)\left(\frac{R_{\rm c}/R_{\rm i}}{30}\right)^{-5/2}.
\end{split}
\end{equation}
This is $\sim$ four orders of magnitude larger than the value quoted in \citet{kasen10} (see their Equation 18). In the most optimistic setting where the opacity drops to $\kappa \sim 0.0005$ cm$^{2}$ g$^{-1}$ after recombination and reaching temperatures $\sim 5\times 10^{3}$ K (see Figure 3 of \citealt{kasen10}) at a time of $R_{\rm c}/R_{\rm i} \simeq 50$, as estimated from Equation \eqref{Tgas}, the optical depth is still $\tau \sim 2\times 10^{4}$ and thus the stream is still very optically thick. 

Because of the extremely high optical depth of the stream, Figure \ref{fig:Lrec} represents an optimistic upper limit of the luminosity from hydrogen recombination, and a better estimate can be obtained by calculating the energy that the stream radiates from a thin layer near its surface. \citet{kasen10} used this approach to estimate the luminosity, but the assumption they made of homologous expansion \citet{kasen10} results in a large overestimate of the emitting area of the stream at the time it reaches a temperature of 5,000 K and a corresponding overestimate of the radiative luminosity; if we assume that the gas radiates from near its surface when it drops to this temperature, then we find a luminosity of
\begin{equation}
\begin{split}
L &= \sigma\pi H L T^4 = \pi \sigma R_{\star}^2\left(\frac{R_{\rm c}}{R_{\rm i}}\right)^{5/2}T^{4} \\
&\simeq10^{37}\left(\frac{R_{\star}}{R_{\odot}}\right)^{2}\left(\frac{R_{\rm c}/R_{\rm i}}{50}\right)^{5/2}\left(\frac{T}{5\times10^{3}\textrm{ K}}\right)^{4}\textrm{ erg s}^{-1}.
\end{split}
\end{equation}
Here we used the fact that the length of the stream $L$ is $L = R_{\star}\left(R_{\rm c}/R_{\rm i}\right)^2$ and the width $H$ is $H = R_{\star}\left(R_{\rm c}/R_{\rm i}\right)^{1/2}$, both of which follow from the exact solution in Section \ref{sec:exact}. This is roughly three orders of magnitude below the estimate given in \citet{kasen10} (see their Equation 22). 

We thus conclude that the transient associated with hydrogen recombination is substantially reduced from the estimate given in Equation \eqref{Lrec} and shown in Figure \ref{fig:Lrec}, likely by at least 2-3 orders of magnitude. Therefore, the detectability of such a feature from a TDE is, unfortunately, highly unlikely.

\subsection{Effect on stream structure}
\label{sec:stream}
Maintaining the assumption that the stream is composed purely of hydrogen and neglecting the occupation of higher electronic states within the atom, the gas-energy equation in the limit that all of the recombination energy is transferred efficiently to the thermal energy of the gas is
\begin{equation}
dE+pdV+\epsilon_{\rm H}Ndx = 0, \label{gasen}
\end{equation}
which states that, in addition to adiabatic expansion, thermal energy $E$ can be lost from the gas through ionization or, as in this case, gained through recombination. If one instead allows for the possibility that a fraction of the gas is composed of Helium and/or metals, then the left-hand side includes a sum over the various ionization potentials of the species and the corresponding ionization fractions. In general we expect the largest contribution to come from Hydrogen, and that while adding in Helium and metal fractions will change the result in detail but not at the order of magnitude level (as including, e.g., Helium, with a mass fraction of 30\%, allows for the presence of the larger recombination energy associated with doubly ionized Helium, but reduces the total recombination energy from Hydrogen by $70\%$), but see \citet{kasen10} for the more general case that accounts for the ionization states of Helium. In Equation \eqref{gasen}, $E$, $V$, and $N$ are the total thermal energy, volume, and conserved baryon number of a fluid element. For a pure hydrogen gas, we have
\begin{equation}
E = \frac{3}{2}\left(1+x\right)NkT, \quad P = \left(1+x\right)nkT = \frac{2}{3}\frac{E}{V}.
\end{equation}
Using these relations in Equation \eqref{gasen} and making a few algebraic rearrangments then gives
\begin{equation}
\frac{\partial}{\partial \tau}\ln\left(\frac{p}{\rho^{5/3}}\right)+\frac{2}{3}\frac{\epsilon_{\rm H}\rho}{m_{\rm H}p}\frac{\partial x}{\partial \tau} = 0. \label{gasenrec}
\end{equation}
This last equation demonstrates that the energy due to recombination increases the entropy of the gas, i.e., if $\partial x/\partial t < 0$ and the number of free electrons decreases, the entropy of the gas increases. 

We can use our background solution and perform a rigorous perturbation analysis with the effects of the entropy due to recombination driving the perturbations. However, it is more illuminating to use the following, approximate method to estimate the effects of recombination on the stream structure: if we assume that the adiabatic, background state (the exact solution derived in Section \ref{sec:exact}, which balances both the tidal term from the SMBH and the equation of hydrostatic balance) is approximately upheld, then the second term in Equation \eqref{gasenrec} is ``known''; we therefore have 
\begin{equation}
p = p_0(s_0)\left(\frac{\rho}{\rho_0}\right)^{5/3}\Delta S(\tau), \label{peqgen}
\end{equation}
where
\begin{equation}
\Delta S = \exp\left[-\frac{2}{3}\frac{\epsilon_{\rm H}\rho_{\rm i}}{m_{\rm H}p_{\rm i}}\int_0^{\tau}e^{2\tau}\frac{\partial x}{\partial \tau}d\tau\right]. \label{DeltaS}
\end{equation}
Here we are calculating the entropy change near the stream axis -- where the density is highest and where most of the mass is -- and hence we set $s_0 = 0$. If we now write $\rho = \rho_{\rm i}H^{-2}e^{-2\tau}g_0(s_0)$ and insert this into Equation \eqref{peqgen} and use the Poisson equation and the equation of hydrostatic balance, then we find
\begin{equation}
H = H_{\rm i}e^{\tau/2}\Delta S^{3/4}.
\end{equation}
If $\Delta S = 1$, then we recover the marginally self-gravitating solution with $H \propto e^{\tau/2}$. From Equation \eqref{mudef}, which shows that $\mu^2 \propto H^{-2}$, we therefore have
\begin{equation}
\mu(t) = \mu_{\rm i}\Delta S^{-3/4}. \label{muoft}
\end{equation}

\begin{figure} 
   \centering
   \includegraphics[width=0.475\textwidth]{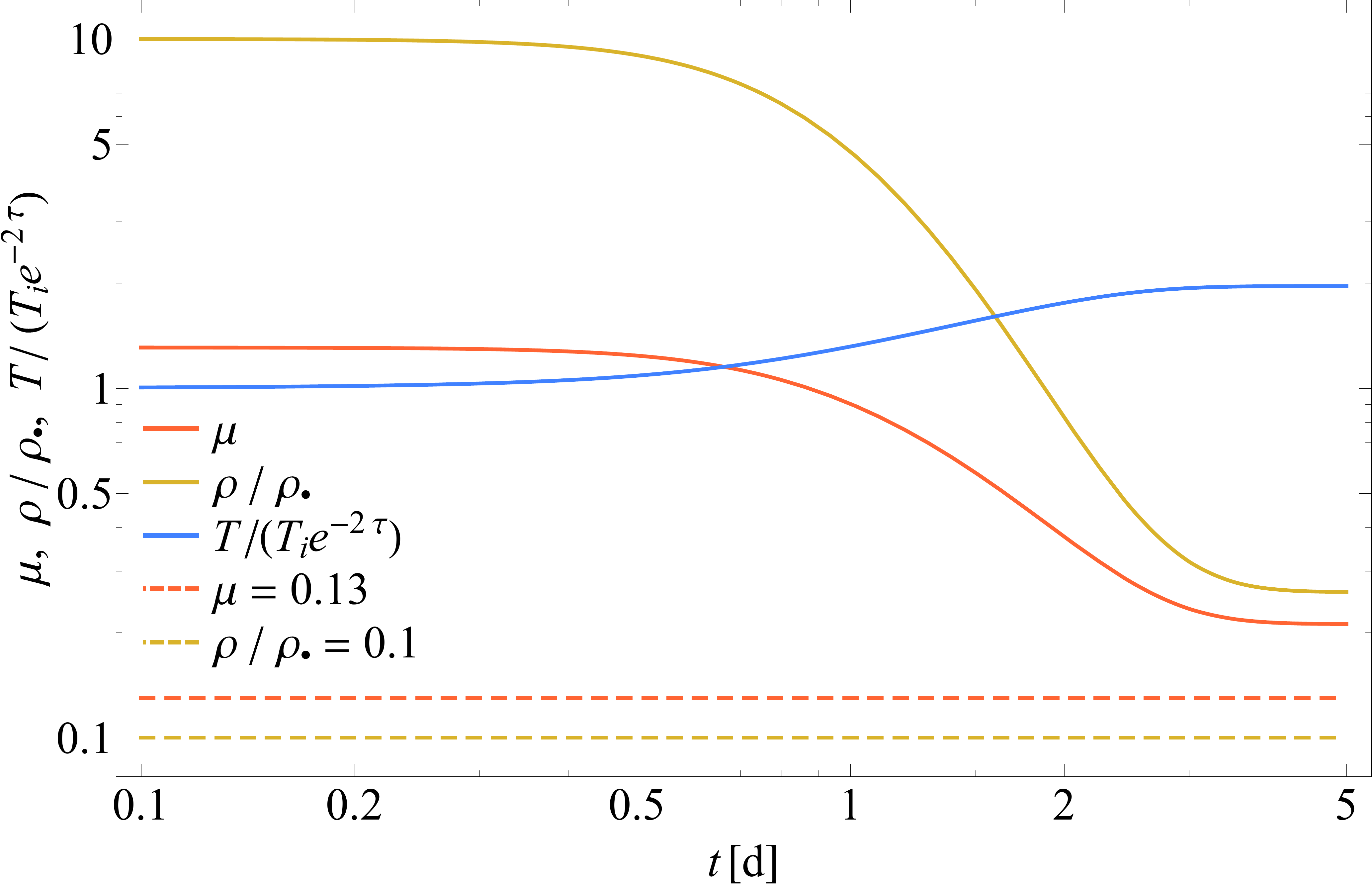} 
   \caption{The value of $\mu \propto \sqrt{\rho/\rho_{\bullet}}$ (orange), the ratio of the stream density to the black hole density $\rho/\rho_{\bullet}$ (yellow), and the temperature relative to the background temperature $T/(T_{\rm i}e^{-2\tau})$ (blue) as a function of time in days for the disruption of a solar-like star. The time dependence of each of these curves is a consequence of recombination within the stream and the entropy deposition therefrom. The horizontal, dashed lines show the critical values of $\mu$ and $\rho/\rho_{\bullet}$ where the stream drops below the self-gravitating limit. }
   \label{fig:muoft}
\end{figure}

Figure \ref{fig:muoft} shows $\mu(\tau)$, $\rho/\rho_{\bullet}$, and $T/(T_{\rm i}e^{-2\tau})$ as a function of time in days with $\Delta S$ calculated from Equation \eqref{DeltaS}, and we assumed the same set of initial conditions as in the previous section (i.e., $\rho_{\rm i} = 10$ g cm$^{-3}$, $T_{\rm i} \simeq 10^{7}$ K). This figure demonstrates that the entropy added to the system as a consequence of recombination substantially lowers the value of $\mu$ and thus the importance of self-gravity. However, the stream asymptotes to a value of $\mu$ that is slightly larger than the minimum value of $\mu_{\rm cr} \simeq 0.13$, and similarly for the ratio of the stream density to the black hole density, meaning that self-gravity is just barely able to confine the stream after all of the recombination energy is transferred to the gas. We also see that the temperature is larger by a factor of $\sim 2$ relative to the background density, and thus the temperature declines less slowly during recombination, as expected.

Equation \eqref{muoft} effectively assumes that the stream starts out in a self-gravitating state (which, as we have argued, should be the case), and as the stream moves from one adiabat (pre-recombination) to another (post-recombination), it remains self-gravitating to the point that the stream width follows from the combination of the Poisson equation and the equation of hydrostatic balance. Figure \ref{fig:muoft} shows that this is at least marginally self-consistent. It also assumes that the fluid velocity imparted by recombination is sufficiently small that the primary contribution to the reduction in the gas density arises from the change in the entropy, and that the temporal derivative of $\mu(\tau)$ can be neglected when computing the eigenvalues of the stream. While the temporal derivative of $\mu$ is exactly zero before and after recombination is complete, as in these two limits the gas is adiabatic (see also Figure \ref{fig:muoft}), the non-zero derivative will complicate the dispersion relation during recombination. Nonetheless, Figure \ref{fig:muoft} shows that $\partial\mu/\partial \tau$ is smooth and occurs over $\gtrsim few$ days, which is many dynamical times of the initial star, and we expect the critical value of $\mu$ derived in the previous section to give a good estimate of when the self-gravitating nature of the stream is destroyed.

In addition to reducing the stream density, recombination must impart a non-zero velocity in the cylindrical-radial direction that would additionally drive the stream away from its self-gravitating state. If $\mu \simeq 1$ initially, then this would lead to the tentative conclusion that hydrogen recombination destroys the stream. We note, however, that our analysis here has ignored the variation in the stream properties along its axis, which should be substantial owing to the increased density in the core relative to the outer extremities of the star. It is likely that self-gravity along the stream increases the density near the marginally bound Keplerian radius, bringing the value of $\mu$ substantially above 1 by the time recombination occurs. 

\begin{figure} 
   \centering
   \includegraphics[width=0.475\textwidth]{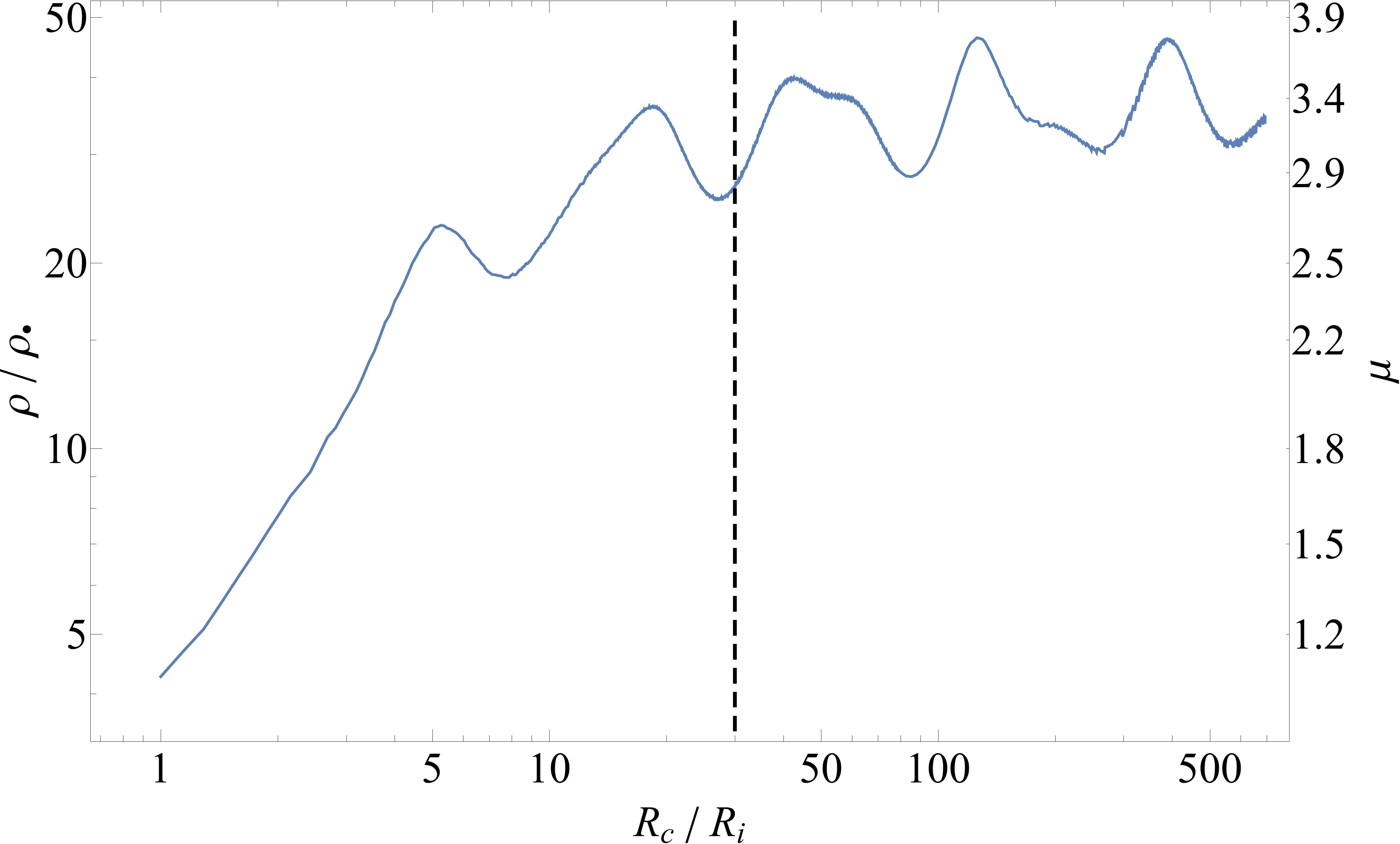} 
   \caption{The ratio of the density at the marginally bound Keplerian radius ($R_{\rm c}$) to the SMBH density as a function of $R_{\rm c}$ that results from a {\sc phantom} simulation of the canonical TDE (a solar-like polytrope disrupted by a $10^6M_{\odot}$ SMBH). The right axis gives the parameter $\mu$, and the vertical lines gives $R_{\rm c}/R_{\rm i} = 30$, at which we expect recombination to heat the gas. }
   \label{fig:rho_rhobullet}
\end{figure}

\begin{figure} 
   \centering
  \includegraphics[width=0.475\textwidth]{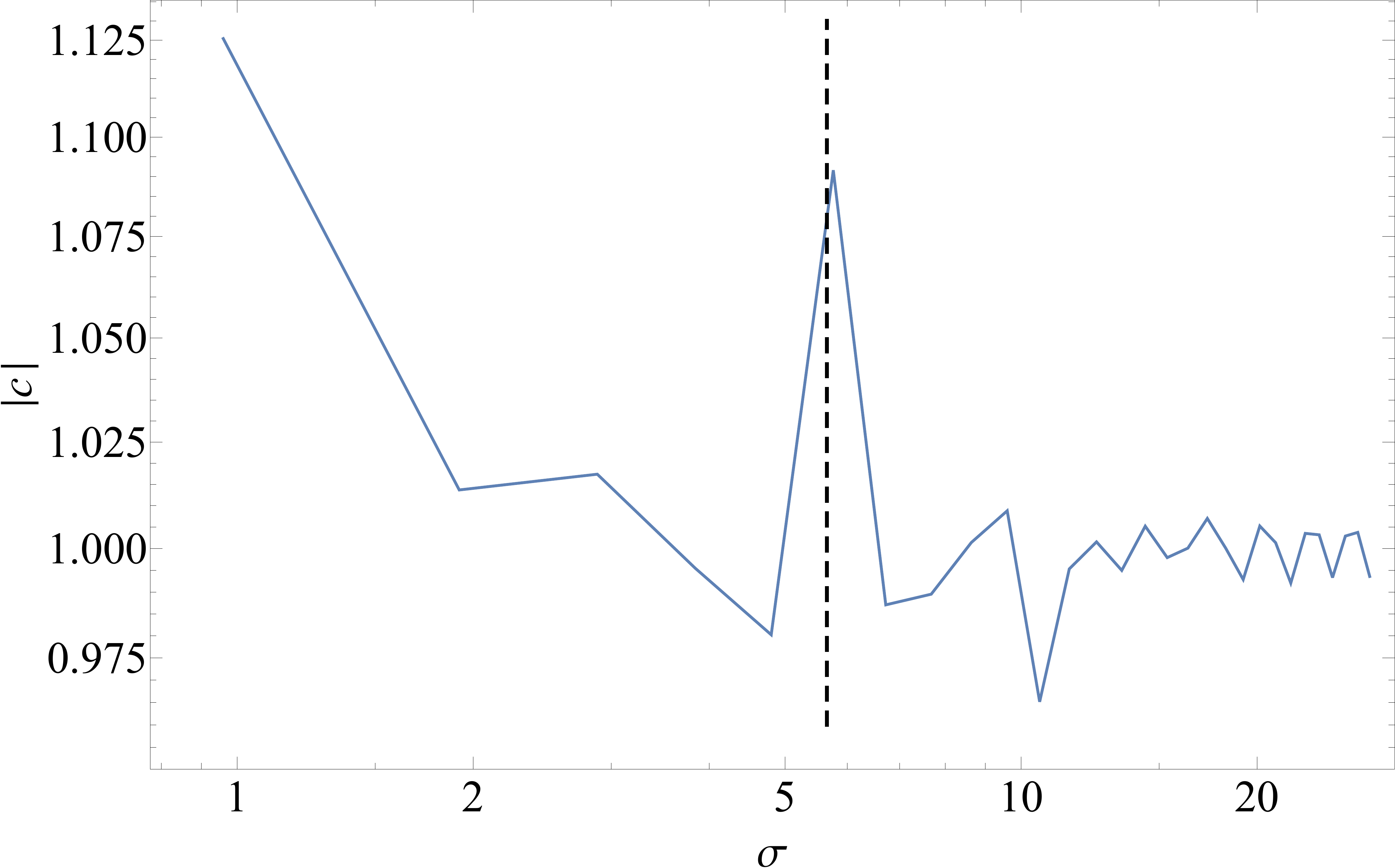} 
   \caption{The temporal power spectrum of $\rho/\rho_{\bullet}$ shown in Figure \ref{fig:rho_rhobullet}; here the coefficients are normalized by the integral under the curve in Figure \ref{fig:rho_rhobullet}, i.e., the temporal average of the density. The vertical, dashed line gives the frequency of the lowest-order oscillatory mode (the ``breathing mode'' of a cylinder) when $\mu = 3$, given by $\sigma \simeq 5.66$, which coincides with the peak exhibited by the power spectrum. }
   \label{fig:fourier_coeff}
\end{figure}

In support of this suggestion, Figure \ref{fig:rho_rhobullet} shows the ratio of the density at the marginally bound Keplerian radius $R_{\rm c}$ to the SMBH density (so the vertical axis is $\propto \rho R_{\rm c}^3$) as a function of the location of the marginally bound radius for the same simulation as in Figures \ref{fig:rho_of_r} and \ref{fig:debris_stream}. The right-hand axis shows the corresponding value of $\mu$. We see that initially the stream density is comparable to the value we expect from the arguments in Section \ref{sec:basic}, but by the time recombination starts to occur vigorously within the stream (shown by the vertical, black dashed line at $R_{\rm c}/R_{\rm i} = 30$), the density (relative to the SMBH density) has increased by an order of magnitude. After this time the ratio $\rho/\rho_{\bullet}$ levels off, but oscillations in the stream are apparent. Figure \ref{fig:fourier_coeff} shows the magnitude of the Fourier coefficients calculated from the Fourier transform of $\rho/\rho_{\bullet}$ in $\tau$, i.e., 

\begin{equation}
c(\sigma) \propto \int e^{i\sigma\tau}\frac{\rho}{\rho_{\bullet}}d\tau.
\end{equation}
We normalized the Fourier coefficients by $c(0)$, i.e., the integral under the curve in Figure \ref{fig:rho_rhobullet}. The vertical, dashed line shows the frequency associated with the ``breathing mode'' of the stream, and that has a value of $\simeq 5.66$ when $\mu = 3$. We see that this frequency coincides almost exactly with the frequency that has noticeably increased power relative to the other frequencies, which strongly suggests that the SPH simulation is capable of resolving the oscillation frequencies of the stream. Nevertheless, we do not see an increase in the amplitude of the perturbation, as would be expected from the fact that the stream is overstable. This is likely due to numerical dissipation, but we leave a detailed investigation of this to future work.

\section{Summary and Conclusions}
\label{sec:summary}
The streams of debris produced from tidal disruption events expand and continue to evolve over many dynamical times of the original, tidally disrupted star, and the question arises as to their stability and self-gravitating nature. We showed that there is an exact solution to the fluid equations (Section \ref{sec:exact}) that describes a self-gravitating, gas-pressure dominated, adiabatic cylinder in the gravitational field of a supermassive black hole, which expands differentially in the cylindrical-radial direction and longitudinally (and at different rates). We identify this solution as the ``background state'' of the gas, and we performed a perturbation analysis of this solution -- considering only cylindrically symmetric perturbations -- in Section \ref{sec:perturbations}. We demonstrated that there is a critical stream density, $\rho_{\rm cr} \simeq 0.1 \rho_{\bullet}$, below which the stream is unstable, and above which the stream is overstable. We identify this density -- which is over an order of magnitude smaller than what is realized in TDEs that is $\rho/\rho_{\bullet} \gtrsim 4$ (see the discussion in Section \ref{sec:basic} and Figures \ref{fig:rho_of_r} and \ref{fig:rho_rhobullet}) -- as the one that divides stable and unstable streams or, more appropriate to the debris streams that are generated from TDEs, regions of a stream that are self-gravitating and those that are not (see Figures \ref{fig:rho_of_r} and \ref{fig:debris_stream}). The growth rate of the overstability is $\propto t^{1/6}$, which is both extremely weakly growing and independent of the stream density. 

We analyzed the effects of hydrogen recombination in Section \ref{sec:recombination}, finding that the stream starts to recombine vigorously by $\sim 80$ dynamical times of the star (or $\sim 1.5$ days for a solar-like star). Because the stream stays very geometrically thin prior to recombination, the density is large and the optical depth across the stream is $\sim 10^{7}$ by the time the stream drops in temperature to $\sim 10^{4}$ K. Consequently, the luminosity of the ``recombination transient'' generated as a fraction of the recombination energy is lost from very near the surface of the stream, is on the order of $\sim 10^{37}$ erg s$^{-1}$ (at most), which is substantially dimmer than what was recovered in previous estimates. We also found that the energy imparted to the gas as a byproduct of recombination was substantial, and caused the density to decline dramatically (see Figure \ref{fig:muoft}). However, because the critical stream density is so far below the density that is typical of TDEs, especially given the tendency of self-gravity to draw material into the denser regions of the stream near the marginally bound Keplerian radius (see Figure \ref{fig:rho_rhobullet}), we find it unlikely that radiative recombination is able to completely destroy the influence of self-gravity. 

It would be interesting to analyze other, physical effects on the debris stream evolution by using the exact solution for the background state. For example, our treatment of the recombination within the stream was highly simplistic, but it would be tenable to use this solution as the hydrodynamic state for a more detailed, radiative transfer calculation. Similarly, it should be possible to analyze the behavior of the magnetic field within the stream, and perhaps even the magnetohydrodynamics, with the exact solution outlined here.

We did not analyze the response of the fluid to perturbations that are along the axis of the stream. These are likely to be important, as not only is the perturbation along the axis of the stream large initially (owing to the density gradient within the star), but these modes -- specifically those that have approximately uniform motion along the stream (i.e., nearly independent of cylindrical radius) -- are gravitationally unstable in the hydrostatic limit (see the analysis and discussion in \citealt{coughlin20} for the case of a polytropic cylinder). We did not analyze these modes here because the exact solution, the cylindrically symmetric perturbations, and the effects of recombination are sufficiently nontrivial and important for the structure of the stream that they deserve discussion in their own right. We will present the analysis of non-cylindrically-symmetric perturbations in a future paper. 

\section*{Data Availability}
The data underlying this article will be shared on reasonable request. 

\section*{Acknowledgements}
I thank Chris Nixon for useful discussions. I thank Chris Kochanek and Dan Kasen for useful correspondence, and the anonymous referee for a careful reading of an initial version of this manuscript and providing comments and suggestions that improved its readability. I acknowledge support from the National Science Foundation through grant AST-2006684 and the Oakridge Associated Universities through a Ralph E.~Powe junior faculty enhancement award.

\bibliographystyle{mnras}

\appendix
\section{Homologous Solutions}
\label{sec:homologous}
The analysis in Section \ref{sec:perturbations} perturbed an exact solution to the fluid equations, which describes a cylindrical and self-gravitating stream expanding in the tidal field of a SMBH (as derived in Section \ref{sec:exact}), to determine the oscillation frequencies and growth rates of small perturbations on top of that exact solution. The oscillation frequencies, commonly referred to as eigenvalues, are dependent on the precise density (and pressure) profile of the unperturbed state, and the eigenfunctions that describe the spatial variation of the fluid as a function of cylindrical radius are similarly functions of that state. The analysis is linear and ignores products of perturbed quantities, and thus breaks down once the unstable modes grow to a level that is comparable to the background solution.

Here we show that an alternative method can be used to analyze the \emph{nonlinear} response of the stream to perturbations, but at the expense of using only the leading-order solution for the density of the background state. In particular, we demonstrate that a simplified variant of the one-zone model of \citet{kochanek94} can be rigorously obtained from the leading-order of an expansion of the current Lagrangian cylindrical radius, $s$, about the original Lagrangian radius, $s_0$, in the fluid equations (this was also recently implemented in \citealt{coughlin22b} to study deep TDEs in which the stellar center of mass comes well within the tidal radius of the SMBH). We assume that the density profile of the stream depends only on cylindrical radius, and that to leading order in cylindrical distance from the stream axis, the initial density profile satisfies
\begin{equation}
\rho_0(s_0, z_0) = \rho_{\rm i}\left\{1-(s_0/\alpha)^2\right\},
\end{equation}
where $\alpha$ is a scale length and $\rho_{\rm i}$ is the density at $s_0 = 0$. This expression must generally hold to leading order in $s_0$ if the density is both well-behaved along the filament axis and cylindrically symmetric. We also assume that there is no dependence of the fluid variables on the distance along the stream axis, $z$; then the solution to Equation \eqref{zmom1} is\footnote{One can generalize this solution to include the decaying branch, $\propto e^{-\tau/2}$, but it has no relevance on the long-term evolution of the stream and so we ignore it here.}
\begin{equation}
z = e^{2\tau}z_0, \quad \tau = \ln\left(\frac{R_{\rm c}}{R_{\rm i}}\right),
\end{equation}
where $R_{\rm i}$ is an arbitrary scale length along the axis of the stream and $R_{\rm c}$ is the marginally bound Keplerian radius that satisfies
\begin{equation}
\frac{\partial R_{\rm c}}{\partial t} = \sqrt{\frac{2GM_{\bullet}}{R_{\rm c}}}.
\end{equation} 
We let the leading-order (in $s_0$) solution for the cylindrical displacement be homologous, i.e., we let
\begin{equation}
s = H(\tau)s_0,
\end{equation}
where $H(0) = 1$ by construction. Then from the Lagrangian solution to the continuity equation \eqref{rhoLag}, the time-dependent density is
\begin{equation}
\begin{split}
\rho &= \left(\frac{s}{s_0}\right)^{-1}\left(\frac{\partial s}{\partial s_0}\right)^{-1}\left(\frac{\partial z}{\partial z_0}\right)^{-1}\rho_0(s_0) \\ 
&=  \rho_{\rm i}e^{-2\tau}H^{-2}\left\{1-\left(s_0/\alpha\right)^2\right\}, \label{rhoA}
\end{split}
\end{equation}
while the pressure follows (to leading order in $s_0$) from the entropy equation \eqref{eosad}:
\begin{equation}
\begin{split}
p &= p_{\rm i}\left(\frac{\rho}{\rho_{\rm i}}\right)^{5/3} \\
&=p_{\rm i}H^{-10/3}e^{-10\tau/3}\left\{1-\frac{5}{3} \left(s_0/\alpha\right)^2\right\}, \label{pA}
\end{split}
\end{equation}
where $p_{\rm i}$ is the pressure along the filament axis. We assumed that the fluid is isentropic (specific and dimensionless entropy function $K \equiv 1$) for concreteness. The solution to the Poisson equation \eqref{poisson} is
\begin{equation}
\frac{\partial \Phi}{\partial s} = 2\pi G\alpha \rho_{\rm i}e^{-2\tau}H^{-1}s_0. \label{PhiA}
\end{equation}
We can now insert Equations \eqref{rhoA} -- \eqref{PhiA} into the $s$-momentum equation \eqref{smom1}, Taylor expand to first order in $s_0$, and change variables from $t$ to $\tau$; note that the latter is
\begin{equation}
\begin{split}
\frac{\partial}{\partial t} &= \frac{\partial\tau}{\partial t}\frac{\partial}{\partial \tau} = \frac{\sqrt{2GM_{\bullet}}}{R_{\rm c}^{3/2}}\frac{\partial}{\partial \tau} = \frac{\sqrt{2GM_{\bullet}}}{R_{\rm i}^{3/2}}e^{-3\tau/2}\frac{\partial}{\partial \tau} \\
\frac{\partial^2}{\partial t^2} &= \frac{2GM_{\bullet}}{R_{\rm i}^{3}}e^{-3\tau/2}\frac{\partial}{\partial \tau}\left[e^{-3\tau/2}\frac{\partial}{\partial \tau}\right] \\ 
&= \frac{GM_{\bullet}}{R_{\rm c}^3}\left(2\frac{\partial^2}{\partial\tau^2}-3\frac{\partial}{\partial \tau}\right).
\end{split}
\end{equation}
Inserting this into the $s$-momentum equation and performing a few lines of algebra, we find that the dynamical equation for $H$ is
\begin{equation}
2\frac{\partial^2H}{\partial \tau^2}-3\frac{\partial H}{\partial \tau}+H+\frac{1}{2}\mu^2\frac{e^{\tau}}{H^2}\left(H-N H^{-1/3}e^{2\tau/3}\right) = 0, \label{Seq}
\end{equation}
where 
\begin{equation}
\frac{2\gamma p_{\rm i}}{\alpha^2 \rho_{\rm i}} \equiv N 2\pi G\rho_{\rm i}, \quad \mu^2 \equiv \frac{4\pi \rho_{\rm i}R_{\rm i}^3}{M_{\bullet}} = 3\frac{\rho_{\rm i}}{\rho_{\bullet}}.
\end{equation}
Here $\rho_{\bullet} = M_{\bullet}/(4\pi R_{\rm i}^3/3)$ and $\mu$ are the same quantities as defined in Section \ref{sec:exact}. 

When $N = 1$, it can be verified that an exact, ``equilibrium'' solution to Equation \eqref{Seq} is
\begin{equation}
H_{\rm eq} = e^{\tau/2}.
\end{equation}
With this solution, the dynamical terms cancel, and the self-gravity and pressure terms balance exactly (i.e., are in equilibrium). We now perturb the solution about this equilibrium by letting
\begin{equation}
H = e^{\tau/2}\left\{1+\delta H(\tau)\right\}, \label{Heq}
\end{equation}
inserting this expression into Equation \eqref{Seq} and keeping first-order terms in $\delta H$. Setting $N = 1$ and linearizing gives
\begin{equation}
2\delta\ddot{H}-\delta\dot{H}+\frac{2\mu^2}{3}\delta H = 0,
\end{equation}
where dots denote differentiation with respect to $\tau$. The solutions to this are $\delta H \propto e^{\sigma\tau}$ with
\begin{equation}
\sigma = \frac{1\pm \sqrt{1-16\mu^2/3}}{4}.
\end{equation}
This demonstrates that when $\mu^2 < \mu_{\rm cr}^2 = 3/16 \simeq 0.188$, or $\mu < \mu_{\rm cr} = \sqrt{3}/4 \simeq 0.433$, the eigenvalues are purely real, one of which is positive and leads to the growth of the perturbations. In the limit that $\mu \rightarrow 0$, the unstable mode has $\sigma = 1/2$, which corresponds to the free expansion of the stream (there is also the solution $\sigma = 0$, which in the perturbative limit represents the freedom to rescale $H$, i.e., in the non-self-gravitating limit it is only the velocity of the fluid that enters into the linearized equation). When $\mu^2 \equiv 3/16$, $\sigma = 1/4$ is a repeated root, and the instability grows as $\propto \tau e^{\tau/4}$. These are the same results that we found in Section \ref{sec:perturbations}, but the value of $\mu^2$ at which the solution goes from purely unstable to overstable is larger by a factor of $\sim 2$, and $\mu_{\rm cr}$ is larger than the exact value by a factor of $\sim 4$. Correspondingly, there is a critical ratio of the stream density to the SMBH density,
\begin{equation}
\frac{\rho_{\rm cr}}{\rho_{\bullet}} = \frac{1}{3}\mu^2 = \frac{1}{16} \simeq 0.063,
\end{equation}
that separates overstable and unstable expansion in the gravitational field of the SMBH. This is a factor of $\sim 2$ smaller than we found in Section \ref{sec:perturbations}.

\begin{figure} 
   \centering
   \includegraphics[width=0.475\textwidth]{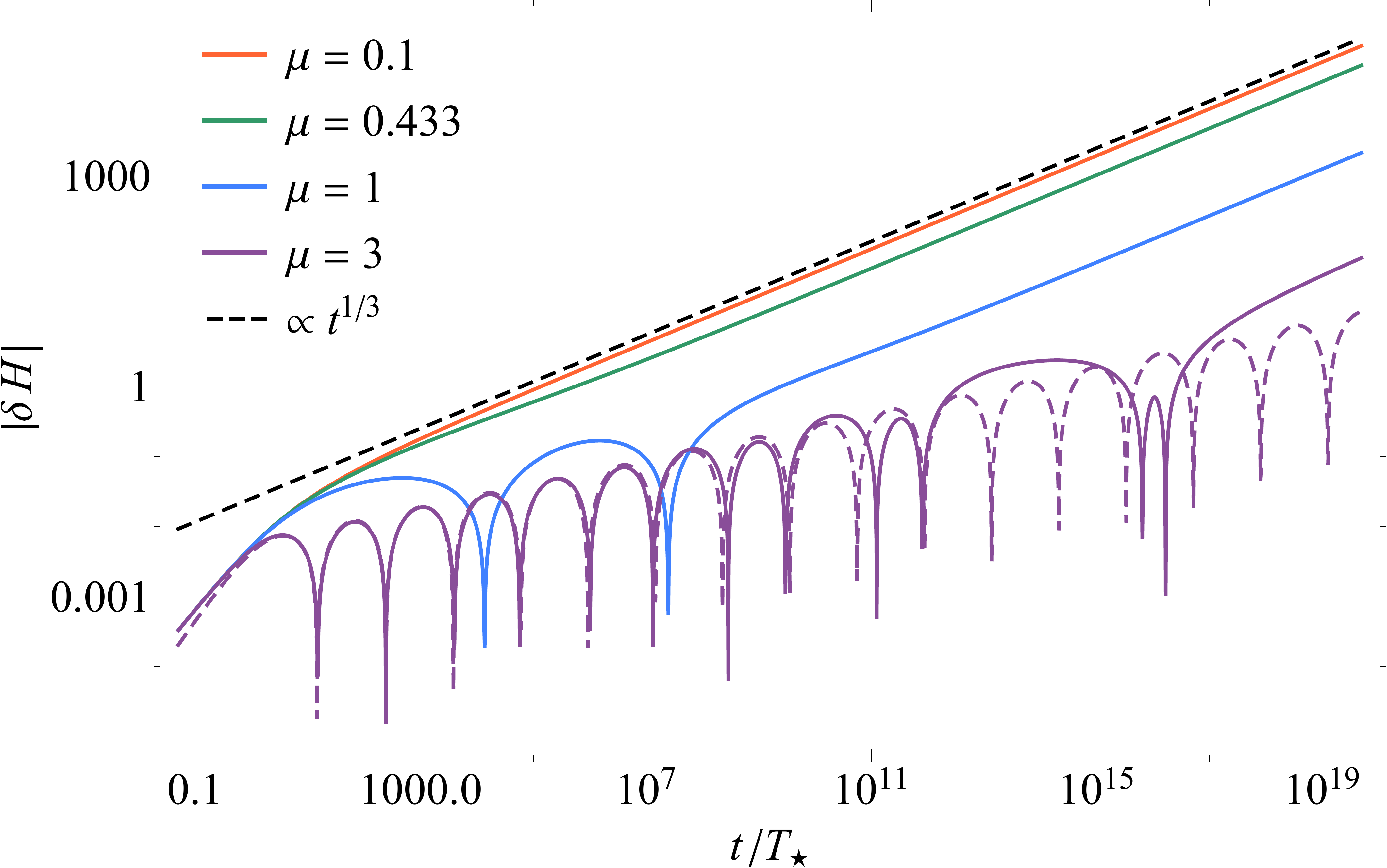} 
   \caption{The absolute value of the correction to the stream width that comes from the solution to Equation \eqref{Seq}, where the $\mu$ for each curve is shown in the legend. Here we set $\delta\dot{H}(0) = 10^{-3}$, which seeds the perturbations initially. The black, dashed curve is $\propto t^{1/3}$, which implies that the stream enters a phase in which it is not bounded by self-gravity, while the dashed, purple curve gives the approximate solution we would expect from linear perturbation theory for $\mu = 3$.}
   \label{fig:deltaH}
\end{figure}

Figure \ref{fig:deltaH} illustrates the absolute value of $\delta H$ that results from Equation \eqref{Seq}, i.e., we solved Equation \eqref{Seq} for $H(\tau)$ and $\delta H \equiv (H-H_{\rm eq})/H_{\rm eq}$, with $H_{\rm eq}$ given in Equation \eqref{Heq}, which is shown in this figure for the $\mu$ given in the legend. Here we seeded the perturbations by letting $\dot\delta{H}(0) = 0.001$, meaning that the stream is only slightly perturbed from its equilibrium solution. We see that for $\mu \le \mu_{\rm cr} \simeq 0.433$, the solution asymptotically approaches $\delta H \propto t^{1/3}$, which implies that the total solution scales as $H \propto e^{2\tau/3} \propto R_{\rm c}$, which is just the solution that one would obtain by neglecting self-gravity. The curves with $\mu = 1$ and $\mu = 3$ oscillate a number of times before they ``bounce'' out of equilibrium and approach the $\propto t^{1/3}$ scaling; when $\mu = 1$ ($\mu = 3$), the solution diverges from the overstable and self-gravitating solution at $t/T_{\star} \simeq 10^{9}$ ($t/T_{\star} \simeq 10^{17}$), i.e., the solutions are effectively stable. The purple, dashed curve in this figure gives the expected variation from linear perturbation theory, which agrees with the numerical solution extremely well until $|\delta H| \simeq 1$, at which point the numerical solution becomes more erratic and the oscillation period less regular.

The solution here only accounts for the first-order terms in the density and density profile about the stream axis. By including higher-order terms, we would recover a limiting $\mu_{\rm cr}$ that is in closer agreement with the solution found in Section \ref{sec:perturbations}, namely $\mu_{\rm cr} \simeq 0.13$. Thus, the extent to which the solutions deviate from the self-gravitating, overstable solutions in Figure \ref{fig:deltaH} is likely overestimated. Including more terms in the series expansion of $s(s_0,\tau)$ would also permit the formation of shocks, which occur because the inversion $s_0(s,\tau)$ is no longer one-to-one, and particles can cross. Shocks likely occur as the stream width rapidly compresses during the overstable phase, as also argued in \citet{kochanek94}, and the large oscillations that grow in amplitude that we recover here are similar to those in Figure 3 of their paper.

\bsp	
\label{lastpage}
\end{document}